\documentclass[11pt]{article}%
\usepackage{amsmath}
\usepackage{adjustbox}
\usepackage{amsfonts}
\usepackage{amsthm}
\usepackage{amssymb}
\usepackage{graphicx}
\usepackage{natbib}
\usepackage{xr-hyper}
\usepackage[colorlinks,citecolor=blue,urlcolor=blue,hyperfootnotes=false]%
{hyperref}
\usepackage[bottom ]{footmisc}
\usepackage[nodisplayskipstretch, onehalfspacing]{setspace}
\usepackage[top=.8in, bottom=.9in, left=0.8in, right=0.8in]{geometry}
\usepackage{caption}
\usepackage[margin=20pt,font+=small,labelformat=parens,labelsep=space,skip=6pt,list=false,hypcap=false]%
{subcaption}
\usepackage{tikz}
\usepackage[sort]{cleveref}
\usepackage{ctable}
\usepackage{afterpage}
\usepackage{framed}
\usepackage{float}
\usepackage[section]{placeins}
\usepackage[english]{babel}
\usepackage[normalem]{ulem}%
\providecommand{\U}[1]{\protect\rule{.1in}{.1in}}
\allowdisplaybreaks
\renewcommand\arraystretch{1.15}
\theoremstyle{plain}
\newtheorem{theorem}{Theorem}

\newtheorem{algorithm}{Algorithm}

\newtheorem{corollary}{Corollary}

\newtheorem{lemma}{Lemma}

\newtheorem{remark}{Remark}

\newcommand{\autocite}[1]{ \textbf{(#1)} }
\newcommand{\textcite}[1]{ \textbf{#1} }

\newtheorem{assumption}{Assumption}

\numberwithin{sassumption}{section}

\crefname{assumption}{Assumption}{Assumptions}
\crefname{sassumption}{Assumption}{Assumptions}
\crefname{theorem}{Theorem}{Theorems}
\crefname{equation}{Equation}{Equations}
\crefname{corollary}{Corollary}{Corollaries}
\crefname{algorithm}{Algorithm}{Algorithms}
\crefname{lemma}{Lemma}{Lemmas}
\crefname{figure}{Figure}{Figures}
\numberwithin{equation}{section}
\begin{document}

\title{Difference-in-Differences with Multiple Time Periods\thanks{First complete
version: March 23, 2018. A previous version of this paper has been circulated
with the title ``Difference-in-Differences with Multiple Time Periods and an
Application on the Minimum Wage and Employment''. We thank the Editor, the
Associate Editor, two anonymous referees, St\'{e}phane Bonhomme, Carol Caetano, Sebastian
Calonico, Xiaohong Chen, Cl\'{e}ment de Chaisemartin, Xavier D'Haultfoeuille,
Bruno Ferman, John Gardner, Andrew Goodman-Bacon, Federico Gutierrez, Sukjin Han, Hugo
Jales, Andrew Johnston, Vishal Kamat, Qi Li, Tong Li, Jason Lindo, Catherine Maclean, Matt Masten, Magne Mogstad, Tom Mroz, Aureo de
Paula, Jonathan Roth, Donald Rubin, Bernhard Schmidpeter, Yuya Sasaki, Na'Ama
Shenhav, Tymon S{\l }oczy\'{n}ski, Sebastian Tello-Trillo, Alex Torgovitsky,
Jeffrey Wooldridge, Haiqing Xu and several seminar and conference audiences
for comments and suggestions. Code to implement the methods proposed in the
paper is available in the R package \texttt{did} which is available on CRAN.}}
\author{Brantly Callaway\thanks{Department of Economics, University of Georgia. Email:
brantly.callaway@uga.edu}
\and Pedro H. C. Sant'Anna\thanks{Department of Economics, Vanderbilt University.
Email: pedro.h.santanna@vanderbilt.edu}}
\date{December 1, 2020}
\maketitle

\begin{abstract}
\onehalfspacing{
In this article, we consider identification, estimation, and inference procedures for treatment effect parameters using Difference-in-Differences (DiD)
with (i) multiple time periods, (ii) variation in treatment timing, and  (iii) when the ``parallel trends assumption" holds
potentially only after conditioning on observed covariates.  We show that a family of causal effect parameters are identified in staggered DiD setups, even if differences in observed characteristics create non-parallel outcome dynamics between groups. Our identification results allow one to use outcome regression, inverse probability weighting, or doubly-robust estimands. We also propose different aggregation schemes that can be used to highlight treatment effect heterogeneity
across different dimensions as well as to summarize the overall effect of participating in the treatment. We establish the asymptotic properties of the proposed estimators and prove the validity of a
computationally convenient bootstrap procedure to conduct asymptotically valid
simultaneous (instead of pointwise) inference.
Finally, we illustrate the relevance of our proposed tools by analyzing the effect of the minimum wage on
teen employment from 2001--2007.  Open-source software is available for implementing the proposed methods.
}

\end{abstract}


\pagebreak

\section{Introduction}

Difference-in-Differences (DiD) has become one of the most popular research
designs used to evaluate causal effects of policy interventions. In its
canonical format, there are two time periods and two groups: in the first
period no one is treated, and in the second period some units are treated (the
treated group), and some units are not (the comparison group). If, in the
absence of treatment, the average outcomes for treated and comparison groups
would have followed parallel paths over time (which is the so-called parallel
trends assumption), one can estimate the average treatment effect for the
treated subpopulation (ATT) by comparing the average change in outcomes
experienced by the treated group to the average change in outcomes experienced
by the comparison group. Methodological extensions of DiD methods often focus
on this standard two periods, two groups setup; see, e.g., \cite{Heckman1997,
Heckman1998}, \cite{Abadie2005}, \cite{Athey2006}, \cite{Qin2008},
\cite{Bonhomme2011}, \cite{DeChaisemartin2017}, \cite{Botosaru2017}, 
\cite{Callaway2018}, and \cite{SantAnna2020}.\footnote{See Section 6 of
\cite{Athey2006} and Theorem S1 in \cite{DeChaisemartin2017} for notable
exceptions that cover multiple periods and multiple groups.}

Many DiD empirical applications, however, deviate from the canonical DiD setup
and have more than two time periods and variation in treatment timing. In this
article, we provide a unified framework for average treatment effects in DiD 
setups with multiple time periods, variation in treatment timing, and when the
parallel trends assumption holds potentially only after conditioning on
observed covariates. We concentrate our attention on DiD with staggered
adoption, i.e., to DiD setups such that once units are treated, they remain
treated in the following periods. 

The core of our proposal relies on separating the DiD analysis into three
separate steps: $\left(  i\right)  $ identification of policy-relevant
disaggregated causal parameters; $\left(  ii\right)  $ aggregation of these
parameters to form summary measures of the causal effects; and $\left(
iii\right)  $ estimation and inference about these different target
parameters. Our approach allows for estimation and inference on interpretable 
causal parameters allowing for arbitrary treatment effect heterogeneity and
dynamic effects, thereby completely avoiding the issues of interpreting
results of standard two-way fixed effects (TWFE) regressions as causal effects
in DiD setups as pointed out by \citet{Borusyak2017},
\citet{DeChaisemartin2016}, \citet{Goodman-bacon2017}, \citet{Abraham2018},
and \citet{Athey2018}. In addition, it adds transparency and objectivity to
the analysis (\citet{Rubin2007, Rubin2008}), and allows researchers to exploit
a variety of estimation methods to answer different questions of interest.

The identification step of the analysis provides a blueprint for the other
steps. In this paper, we pay particular attention to the disaggregated causal
parameter that we call the \textit{group-time average treatment effect}, i.e.,
the average treatment effect for group $g$ at time $t$, where a
\textquotedblleft group\textquotedblright\ is defined by the time period when
units are first treated. In the canonical DiD setup with two periods and two
groups, these parameters reduce to the ATT which is typically the parameter of
interest in that setup. An attractive feature of the group-time average
treatment effect parameters is that they do not directly restrict
heterogeneity with respect to observed covariates, the period in which units are first treated, or the evolution of treatment effects over time. As a
consequence, these easy-to-interpret causal parameters can be directly used
for learning about treatment effect heterogeneity, and/or to construct many
other more aggregated causal parameters. We view this level of generality and
flexibility as one of the main advantages of our proposal.

We provide sufficient conditions related to treatment anticipation behavior
and conditional parallel trends under which these group-time average treatment
effects are nonparametrically point-identified. A unique feature of our
framework is that it shows how researchers can flexibly incorporate covariates
into the staggered DiD setup with multiple groups and multiple periods. This
is particularly important in applications in which differences in observed
characteristics create non-parallel outcome dynamics between different groups -- in this case, unconditional DiD strategies are generally not appropriate to 
recover sensible causal parameters of interest
\citep{Heckman1997, Heckman1998,Abadie2005}. We propose three different types
of DiD estimands in staggered treatment adoption setups: one based on outcome
regressions \citep{Heckman1997, Heckman1998}, one based on inverse probability
weighting \citep{Abadie2005}, and one based on doubly-robust methods
\citep{SantAnna2020}. We provide versions of these estimands both for the case
with panel data and for the case with repeated cross sections data. To the
best of our knowledge, this paper is the first to show how one can allow for
covariate-specific trends across groups in DiD setups with variation in
treatment timing. Our results also highlight that, in practice, one can rely
on different types of parallel trends assumptions and allow some types of
treatment anticipation behavior; our proposed estimands explicitly reflect
these assumptions.

Our framework acknowledges that in some applications there may be many
group-time average treatment effects and researchers may want to aggregate
them into different summary causal effect measures. This characterizes the
aggregation step of the analysis. We provide ways to aggregate the potentially
large number of group-time average treatment effects into a variety of
intuitive summary parameters and discuss specific aggregation schemes that can
be used to highlight different sources of treatment effect heterogeneity
across groups and time periods. In particular, we consider aggregation schemes
that deliver a single overall treatment effect parameter with similarities to
the ATT in the two period and two group case as well as partial aggregations
that highlight heterogeneity along certain dimensions such as $\left(
a\right)  $ how average treatment effects vary with length of exposure to the
treatment (event-study-type estimands); $\left(  b\right)  $ how average
treatment effects vary across treatment groups; and $\left(  c\right)  $ how
cumulative average treatment effects evolve over calendar time. We also
provide a formal discussion of the costs and benefits of balancing the sample
in \textquotedblleft event time\textquotedblright\ when analyzing dynamic
treatment effects. Overall, our setup makes it clear that, in general, the
\textquotedblleft best\textquotedblright\ aggregation scheme is
application-specific as it depends on the type of question one wants to answer.

Given that our identification results are constructive, we propose easy-to-use
plug-in type (parametric) estimators for the causal parameters of interest.
Although the outcome regression, inverse probability weighting and
doubly-robust estimands are equivalent from the identification point of view,
they suggest different types of DiD estimators one can use in practice. Here,
we note that using doubly-robust estimators can be particularly attractive as
they rely on less stringent modeling conditions than the outcome regression
and the inverse probability weighting procedures.

In order to conduct asymptotically valid inference, we justify the use of a
computationally convenient multiplier-type bootstrap procedure. This approach
can be used to obtain simultaneous confidence bands for the group-time average
treatment effects. Unlike commonly used pointwise confidence bands, our
simultaneous confidence bands asymptotically cover the \textit{entire path} of
the group-time average treatment effects with {fixed probability} and take
into account the dependency across different group-time average treatment
effect estimators. Thus, our proposed confidence bands are arguably more
suitable for visualizing the overall estimation uncertainty than more
traditional pointwise confidence intervals.

We illustrate the practical relevance of our proposal by analyzing the effect
of the minimum wage on teen employment. Here, we follow much empirical work on
the effects of the minimum wage and exploit having access to panel data and
variation in treatment timing across states (e.g.,
\cite{Card1994,Neumark2000,Neumark2008,Dube2010}, among many others) in order
to estimate the effect of the minimum wage on employment. Interestingly, in
our setup, using our approach leads to qualitatively different results than
results from the TWFE estimator. This suggests that, at least in certain
applications, using methods that are robust to treatment effect heterogeneity
can lead to meaningful differences relative to more standard TWFE
regressions.\bigskip

\textbf{Recent Related Literature: }This paper is related to the recent and
emerging literature on heterogeneous treatment effects in DiD and/or event
studies with variation in treatment timing; see, e.g.,
\cite{DeChaisemartin2016}, \cite{Goodman-bacon2017}, \cite{Imai2018},
\cite{Borusyak2017}, \cite{Athey2018} and \cite{Abraham2018}. All these papers
present, among other things, some negative\ results about the interpretation
of parameters associated with standard TWFE linear regression specifications;
see also \cite{Laporte2005}, \cite{Wooldridge2005}, \cite{Chernozhukov2013},
and \cite{Gibbons2018} for earlier related results based on (one-way)
fixed-effect estimators. Our proposed procedure completely bypasses the
pitfalls highlighted in these papers as we clearly separate the
identification, aggregation and estimation/inference steps of the analysis.

These aforementioned papers also propose alternative DiD estimators that do
not suffer from the pitfalls associated with TWFE. Among these, perhaps
the closest to our proposal are those of \cite{DeChaisemartin2016}, and
\cite{Abraham2018}, though several major differences are worth stressing.

\cite{DeChaisemartin2016} is focused on recovering an instantaneous
treatment effect measure, while we pay particular attention to treatment
effect dynamics. In fact, our framework allows one to form \textit{families}
of different aggregate parameters in a unified manner. Second, while we pay
particular attention to the role played by pre-treatment covariates,
\cite{DeChaisemartin2016} mainly focus on unconditional DiD 
designs. On the
other hand, the setup in \cite{DeChaisemartin2016} is more general than ours
as we consider staggered adoption designs and they allow for more general
treatment selection. Nonetheless, we note that the unconditional versions of
our parallel trends assumptions are weaker than the one in
\cite{DeChaisemartin2016}, even if one were to specialize their setup to
staggered adoption designs.

\cite{Abraham2018} proposes a parameter, cohort-specific average treatment
effects, that translates our group-time average treatment effects from
calendar time into event time. \cite{Abraham2018} proposes regression-based
estimators of these parameters that have similar properties to our estimators
in the specific case of staggered treatment adoption under an unconditional
version of the parallel trends assumption. However, our approach is more
general in several respects. First, we allow for parallel trends assumptions
to hold after conditioning on covariates, and it is not clear how to adapt the
regression based estimators in \cite{Abraham2018} to this case. Second, we
consider a wide variety of possible aggregations of group-time average
treatment effects where \cite{Abraham2018} focuses particularly on the event
study type of aggregation. Third, we make use of simultaneous inference
procedures that explicitly account for potential multiple-testing problems;
\cite{Abraham2018} focuses on pointwise inference. On the other hand, we do
not have any results highlighting the pitfalls associated with using TWFE
specifications with leads and lags of treatment indicators to conduct causal
inference; these are unique to \cite{Abraham2018}.


We also note that \cite{Athey2018} considers a staggered treatment adoption
setup similar to ours. However, the starting point of \cite{Athey2018} is an
assumption that the treatment adoption date is fully randomized which is
stronger than our parallel trends assumptions. We also note that
\cite{Athey2018} abstracts away from the important role played by covariates
in the DiD analysis and does not consider aggregation schemes to summarize
treatment effect heterogeneity like we do. On the other hand, we stress that
the main focus of their paper is on providing design-based inference
procedures for staggered DiD setups with random treatment dates. Their
design-based inference procedures complement our sampling-based inference
procedures. \bigskip

\textbf{Organization of the paper: }The remainder of this article is organized
as follows. Section 2 presents our main identification results. We discuss our
different aggregation schemes in Section 3. Estimation and inference
procedures for the treatment effects of interest are presented in Section 4.
We revisit the effect of minimum wage on employment in Section 5. Section 6
concludes. Proofs as well as additional methodological results are reported in
the Appendix. In the Supplementary Appendix, we present proofs for the results when only repeated cross-sections data is available, provide additional details about the empirical application, and present a small scale Monte Carlo simulation to illustrate the finite sample properties of our proposed estimators.\footnote{Supplementary Appendix is available at \url{https://pedrohcgs.github.io/files/Callaway_SantAnna_2020_supp.pdf}.}

\section{Identification\label{sec:2}}

\subsection{Setup}

We first introduce the notation we use throughout the article. We consider the
case with $\mathcal{T}$ periods and denote a particular time period by $t$
where $t=1,\ldots,\mathcal{T}$. In a canonical DiD setup, $\mathcal{T}=2$ and
no one is treated in period $t=1$. Let $D_{i,t}$ be a binary variable equal to
one if unit $i$ is treated in period $t$ and equal to zero otherwise. We make
the following assumption about the treatment process:

\begin{assumption}
[Irreversibility of Treatment]\label{ass:irrev} $D_{1}=0$ $almost$ $surely$
(a.s.). For $t=2,\ldots,\mathcal{T}$,
\[
D_{t-1}=1\text{ implies that }D_{t}=1\text{ a.s..}%
\]

\end{assumption}

Assumption \ref{ass:irrev} states that no one is treated at time $t=1,$ and
that once a unit becomes treated, that unit will remain treated in the next
period.\footnote{{In applications, it can be the case that some units are
already treated by the first time period. In our case, we would drop these
units; this is analogous to the case with two time periods. The reason to drop
these units is that untreated potential outcomes are never observed for this
group which will imply that treatment effects are not identified for this group nor 
are they useful as a comparison group under a parallel trends assumption.}}
This assumption is also called staggered treatment adoption in the literature.
We interpret this assumption as if units do not \textquotedblleft
forget\textquotedblright\ about the treatment experience.\footnote{See
\cite{Han2019}, \cite{DeChaisemartin2016} and \cite{Bojinov2020} for
alternative setups where treatment can \textquotedblleft turn
off\textquotedblright.}

Define $G$ as the time period when a unit first becomes treated. Under
\Cref{ass:irrev}, for all units that eventually participate in the treatment,
$G$ defines which \textquotedblleft group\textquotedblright\ they belong to.
If a unit does not participate in any time period, we arbitrarily set
$G=\infty$. We define $G_{g}$ to be a binary variable that is equal to one if
a unit is first treated in period $g$ (i.e., $G_{i,g}=\mathbf{1}\{G_{i}=g\}$)
and define $C$ to be a binary variable that is equal to one for units that do
not {participate in the treatment in any time period} (i.e., $C_{i}%
=\mathbf{1}\{G_{i}=\infty\}=1-D_{i,\mathcal{T}}$). Let $\bar{g}=\max_{i=1,\cdots,n} G_i$ be the maximum $G$ in the dataset. Next, denote the
generalized propensity score as $p_{g,s}(X)=P(G_{g}=1|X,G_{g}+\left(
1-D_{s}\right)  \left(  1-G_{g}\right)  =1)$. Note that $p_{g,s}(X)$ indicates
the probability of being first treated at time $g$, conditional on
pre-treatment covariates $X$ and on either being a member of group $g$ {(in
this case, $G_{g}=1$)} or a member of the \textquotedblleft
not-yet-treated\textquotedblright\ group by time $s$ {(in this case,
$(1-D_{s})(1-G_{g})=1$)}. {Many of our results use a specialized version of
this generalized propensity score, and, henceforth,} we define $p_{g}%
(X)=p_{g,\mathcal{T}}(X)=P(G_{g}=1|X,G_{g}+C=1)$ {which is the probability of
being first treated in period $g$ conditional on covariates and either being a
member of group $g$ or not participating in the treatment in any time period}.
Let $\mathcal{G}=\operatorname*{supp}(G)\backslash\left\{ \bar{g}\right\}
\subseteq\left\{  2,3,\dots,\mathcal{T}\right\}  $ denote the support of $G$
excluding $\bar{g}$.\footnote{When there is a ``never treated'' set of units with $G=\infty$, $\mathcal{G}$ only excludes this group. When such ``never-treated'' group is not available, we exclude the latest-treated group as there will be no valid untreated comparison group for them.} Likewise, let $\mathcal{X}=\operatorname*{supp}%
(X)\subseteq\mathbb{R}^{k}$ denote the support of the pre-treatment
covariates. Finally, for a generic $\delta\geq0$, let $\mathcal{G}_{\delta
}=\mathcal{G\cap}\left\{  2+\delta,3+\delta,\dots,\mathcal{T}\right\}  $.  

Next, we set up the potential outcomes framework. Here, we combine the dynamic
potential outcomes framework of \cite{Robins1986, Robins1987} with the
multi-stage treatment adoption setup discussed by \cite{Heckman2016}; see also
\cite{Sianesi2004}. Let $Y_{i,t}(0)$ denote unit $i$'s untreated potential
outcome at time $t$ if they {remain untreated through time period}
$\mathcal{T}${; i.e., if they were not to participate in the treatment across all
available time periods}. For $g=2,\dots,\mathcal{T}$, let $Y_{i,t}(g)$ denote
the potential outcome that unit $i$ would experience at time $t$ if they were
to first become treated in time period $g$. Note that our potential outcomes
notation accounts for potential dynamic treatment selection, though it also
accommodates (pre-specified) treatment regimes \citep{Murphy2001, Murphy2003}.
The observed and potential outcomes for each unit $i$ are related through%
\begin{equation}
Y_{i,t}=Y_{i,t}\left(  0\right)  +\sum_{g=2}^{\mathcal{T}}\left(
Y_{i,t}\left(  g\right)  -Y_{i,t}\left(  0\right)  \right)  \cdot G_{i,g}
\label{eq:observational}%
\end{equation}
In other words, we only observe one potential outcome path for each unit. For
those that do not participate in the treatment in any time period, observed
outcomes are untreated potential outcomes in all periods. For units that do
participate in the treatment, observed outcomes are the unit-specific
potential outcomes corresponding to the particular time period when that unit
adopts the treatment.

We also impose the following random sampling assumption.

\begin{assumption}
[Random Sampling]\label{ass:iid} $\{Y_{i,1},Y_{,i2},\ldots Y_{i,\mathcal{T}%
},X_{i},D_{i,1},D_{i,2},\ldots,D_{i,\mathcal{T}}\}_{i=1}^{n}$ is independent
and identically distributed $(iid)$.
\end{assumption}

Assumption \ref{ass:iid} implies that we have access to panel data; {our
results extend essentially immediately to the case with repeated cross
sections data and this case is developed in Appendix B.} Here, we note that
Assumption \ref{ass:iid} allows us to view all potential outcomes as random.
Furthermore, it does not impose restrictions between
\emph{potential outcomes} and treatment allocation, nor does it restrict the time series dependence of
the observed random variables. On the other hand, Assumption \ref{ass:iid}
imposes that each unit $i$ is randomly drawn from a large population of
interest. For an alternative design-based inference approach, see
\cite{Athey2018}.

Henceforth, to keep the notation more concise, we will suppress the unit index
$i$ in our notation.

\subsection{The Group-Time Average Treatment Effect Parameter}

Given that different potential outcomes cannot be observed for the same unit
at the same time, researchers often focus on identifying and estimating some
average causal effects. For instance, in the canonical DiD setup {with two
time periods}, the most popular treatment effect parameter of interest is the
average treatment effect on the treated, denoted by\footnote{Existence of
expectations is assumed throughout.}
\[
ATT=\mathbb{E}[Y_{2}(2)-Y_{2}(0)|G_{2}=1].
\]
In this paper, we consider a natural generalization of the $ATT$ that is
suitable to setups with multiple treatment groups and multiple time periods.
More precisely, we use the average treatment effect for units who are members
of a particular group $g$ at a particular time period $t$, denoted by
\[
ATT(g,t)=\mathbb{E}[Y_{t}(g)-Y_{t}(0)|G_{g}=1],
\]
as the main building block of our framework. We call this causal parameter the
\textit{group-time average treatment effect}.

Note that the $ATT\left(  g,t\right)  $ does not impose any restriction on
treatment effect heterogeneity across groups or across time. Thus, focusing on
the family of $ATT(g,t)$'s allow us to analyze how average treatment effects
vary across different dimensions in a unified manner. For instance, by
fixing a group $g$ and varying time $t$, one is able to highlight how average
treatment effects evolve over time for that specific group. By doing this for
different groups, we can have a better understanding about how treatment
effect dynamics vary across groups. In addition, as we discuss in Section
\ref{sec:4}, one can build on the $ATT(g,t)$'s to form more aggregated causal
parameters that are constructed to answer specific questions like: $\left(
a\right)  $ What was the average effect of participating in the treatment
across all groups that participated in the treatment by time period
$\mathcal{T}$? $\left(  b\right)  $ Are average treatment effects
heterogeneous across groups? $\left(  c\right)  $ How do average treatment
effects vary by length of exposure to the treatment? $\left(  d\right)  $
How do cumulative average treatment effects evolve over calendar time? We view
this level of generality and flexibility as one of the main advantages of our
framework that first focuses on the family of $ATT(g,t)$'s.

\subsection{Identifying Assumptions}

In order to identify the $ATT(g,t)$ and their functionals, we impose the
following assumptions.

\begin{assumption}
[Limited Treatment Anticipation]\label{ass:anticipation} There is a known
$\delta\geq0$ such that
\[
\mathbb{E}[Y_{t}(g)|X,G_{g}=1]=\mathbb{E}[Y_{t}(0)|X,G_{g}=1]~a.s.\text{ for
all }g\in\mathcal{G},t\in\left\{  1,\dots,\mathcal{T}\right\}  \text{ such
that }t<g-\delta.
\]

\end{assumption}

Assumption \ref{ass:anticipation} restricts anticipation of the treatment for
all \textquotedblleft eventually treated\textquotedblright\ groups. When
$\delta=0$, it imposes a \textquotedblleft no-anticipation\textquotedblright%
\ assumption, see, e.g., \cite{Abbring2003} and \cite{Sianesi2004}. This is
likely to be the case when the treatment path is not a priori known and/or
when units are not the ones who \textquotedblleft choose\textquotedblright%
\ treatment status. However, Assumption \ref{ass:anticipation} also allows for
anticipation behavior, as long as we have a good understanding about the
anticipation horizon $\delta$. For instance, if units anticipate treatment by
one period, Assumption \ref{ass:anticipation} would hold with $\delta=1$; see,
e.g., \cite{Laporte2005} and \cite{Malani2015} for the importance of
accounting for potential anticipation behavior. Note that, under Assumption
\ref{ass:anticipation}, $ATT\left(  g,t\right)  =0$ for all pre-treatment periods such that $t<g-\delta$.

Next, we consider two alternative assumptions that impose restrictions on the
evolution of untreated potential outcomes.

\begin{assumption}
[Conditional Parallel Trends based on a \textquotedblleft
Never-Treated\textquotedblright\ Group]\label{ass:common-trends} Let $\delta$
be as defined in Assumption \ref{ass:anticipation}. For each $g\in\mathcal{G}$ and $t\in\left\{
2,\dots,\mathcal{T}\right\} $ such that $t\geq g-\delta$,
\[
\mathbb{E}[Y_{t}(0)-Y_{t-1}(0)|X,G_{g}=1]=\mathbb{E}[Y_{t}(0)-Y_{t-1}%
(0)|X,C=1]~a.s..
\]

\end{assumption}

\begin{assumption}
[Conditional Parallel Trends based on \textquotedblleft
Not-Yet-Treated\textquotedblright\ Groups]\label{ass:common-trends-ny} Let
$\delta$ be as defined in Assumption \ref{ass:anticipation}. For each $g\in\mathcal{G}$ and each $\left(
s,t\right)  \in\left\{  2,\dots,\mathcal{T}\right\}  \times\left\{
2,\dots,\mathcal{T}\right\}  $  such that $  t \ge g-\delta $ and
$ t+\delta \le s < \bar{g}$,
\[
\mathbb{E}[Y_{t}(0)-Y_{t-1}(0)|X,G_{g}=1]=\mathbb{E}[Y_{t}(0)-Y_{t-1}%
(0)|X,D_{s}=0,G_{g}=0]~a.s..
\]

\end{assumption}

Assumptions \ref{ass:common-trends} and \ref{ass:common-trends-ny} are two
different conditional parallel trends assumptions that generalize the
two-period parallel trends assumption to the case where there are multiple time periods
and multiple treatment groups; see, e.g., \cite{Heckman1997, Heckman1998},
\cite{Abadie2005} and \cite{SantAnna2020}. Both assumptions hold after
conditioning on covariates $X$. This can be important in many applications in
economics particularly in cases where there are covariate specific trends in
outcomes over time and when the distribution of covariates is different across
groups. For example, \cite{Heckman1997} motivates conditional parallel trends
assumptions in the context of evaluating a job training program. For
evaluating job training programs, the distribution of observed covariates such
as age, employment history, and years of education is often quite different
between individuals who participate in job training and those that do not.
When the path of labor market outcomes (in the absence of participating in job
training) depends on these covariates, a conditional parallel trends becomes more plausible than an unconditional parallel trends assumption.
In fact, ignoring the presence of covariate-specific trends can result in
important biases when evaluating causal effects of policy interventions using
unconditional DiD methods.

Assumptions \ref{ass:common-trends} and \ref{ass:common-trends-ny} differ from
each other depending on the comparison group one is willing to use in a given
application. More specifically, Assumption \ref{ass:common-trends} states
that, conditional on covariates, the average outcomes for the group first
treated in period $g$ and for the \textquotedblleft
never-treated\textquotedblright\ group would have followed parallel paths in
the absence of treatment. Assumption \ref{ass:common-trends-ny} imposes 
conditional parallel trends between group $g$ and groups that are \textquotedblleft
not-yet-treated\textquotedblright\ by time $t+\delta$.\footnote{ \cite{Athey2006} and \cite{DeChaisemartin2017} also consider using ``not-yet-treated'' units as comparison groups in related DiD procedures.} Importantly,
both of these assumptions allow for covariate-specific trends and do not
restrict the relationship between treatment timing and the potential outcomes, 
$Y_{t}\left(  g\right)  $'s. Thus, they are weaker than the
randomization-based assumption made by \cite{Athey2018}. We also note that the
unconditional versions of Assumptions \ref{ass:common-trends} and
\ref{ass:common-trends-ny} are weaker than the parallel trends assumption
imposed by \cite{DeChaisemartin2016} and \cite{Abraham2018} as they impose
fewer restrictions on the evolution of $Y_{t}\left(  0\right)  $ in
pre-treatment periods; see, e.g., \cite{Marcus2020} for a comparison.

In our view, practitioners may favor Assumption \ref{ass:common-trends} with
respect to Assumption \ref{ass:common-trends-ny} when there is a sizable group
of units that do not participate in the treatment in any period, and, at the
same time, these units are similar enough to the \textquotedblleft eventually
treated\textquotedblright\ units. When a \textquotedblleft
never-treated\textquotedblright\ group of units is not available or
\textquotedblleft too small\textquotedblright, researchers may favor
Assumption \ref{ass:common-trends-ny} as it allows one to use more groups as
valid comparison units, which potentially leads to more informative inference
procedures. However, it is important to stress that favoring Assumption
\ref{ass:common-trends-ny} with respect to Assumption \ref{ass:common-trends}
also involves potential drawbacks. For instance, in the absence of treatment
anticipation ($\delta=0$), Assumption \ref{ass:common-trends} does not
restrict observed pre-treatment trends across groups, whereas Assumption
\ref{ass:common-trends-ny} does; see, e.g., \cite{Marcus2020}. Not restricting
pre-treatment trends may be particularly meaningful in applications where the
economic environment during the ``early-periods'' was potentially different
from the ``later-periods.'' In these cases, the outcomes of different groups
may evolve in a non-parallel manner during ``early-periods'', perhaps because
the groups were exposed to different shocks, while trends become parallel in
the ``later-periods.'' We recommend taking these trade-offs into account when
deciding which conditional parallel trends assumption is more appropriate for
a given application.\footnote{It may be tempting to use statistical pre-tests
to select between different versions of the parallel trends assumption.
However, the results of \cite{Roth2020} show that such a practice can lead to
important distortions when conducting inference. Thus, we do not recommend
following this path, but instead recommend taking the context of the
application into account in order to choose the appropriate parallel trends
assumption.}

The final identifying assumption we impose is an overlap condition.

\begin{assumption}
[Overlap]\label{ass:overlap} For each $t\in\left\{  2,\dots,\mathcal{T}%
\right\}  $, $g\in\mathcal{G}$, there exist some $\varepsilon>0$ such that
$P\left(  G_{g}=1\right)  >\varepsilon$ and $p_{g,t}(X)<1-\varepsilon$ $a.s.$.
\end{assumption}

Assumption \ref{ass:overlap} extends the overlap assumption in
\cite{Heckman1997, Heckman1998}, \cite{Abadie2005}, and \cite{SantAnna2020} to
the multiple groups and multiple periods setup. It states that a positive
fraction of the population starts treatment in period $g,$ and that, for all
$g$ and $t$, the generalized propensity score is uniformly bounded away from
one. Assumption \ref{ass:overlap} rules out \textquotedblleft irregular
identification\textquotedblright, see, e.g., \cite{Khan2010}.

\begin{remark}
\label{Rem:anticip}Note that Assumption \ref{ass:anticipation} and Assumption
\ref{ass:common-trends} (Assumption \ref{ass:common-trends-ny}) are
intrinsically connected. For instance, when one imposes the \textquotedblleft
no-anticipation\textquotedblright\ condition (so that $\delta=0$), Assumption
\ref{ass:common-trends} would then impose conditional parallel trends only for
post-treatment periods $t\geq g$. If one allows for anticipation behavior (so
that $\delta>0$), Assumption \ref{ass:common-trends} would then impose
conditional parallel trends in some pre-treatment periods, too. In fact, the
parallel trends assumptions become stronger as one increases $\delta$. To the
best of our knowledge, this trade-off between the strength of these
assumptions has not been noticed before.
\end{remark}

\begin{remark}
In some applications, practitioners may not be comfortable with using
\textquotedblleft never-treated\textquotedblright\ units as part of the
comparison group because they behave very differently from the other
\textquotedblleft eventually treated\textquotedblright\ units. In these cases,
practitioners could drop all \textquotedblleft never-treated\textquotedblright%
\ units from the analysis and proceed with Assumption
\ref{ass:common-trends-ny}.
\end{remark}

\subsection{Nonparametric Identification of the Group-Time Average Treatment
Effects}

In this section, we show that the family of group-time average treatment
effects are nonparametrically point-identified under the aforementioned
assumptions. Furthermore, we show that one can use outcome regression (OR),
inverse probability weighting (IPW), or doubly robust (DR) estimands to
recover the $ATT\left(  g,t\right)  $'s. In addition, we also highlight the
roles played by Assumption \ref{ass:anticipation} and by Assumptions
\ref{ass:common-trends} and \ref{ass:common-trends-ny} when forming these
different estimands.

Before formalizing all the results, we need to introduce some additional
notation. Let $m_{g,t,\delta}^{nev}\left(  X\right)  =\mathbb{E}\left[
Y_{t}-Y_{g-\delta-1}|X,C=1\right]  $ and $m_{g,t,\delta}^{ny}\left(  X\right)
=\mathbb{E}\left[  Y_{t}-Y_{g-\delta-1}|X,D_{t+\delta}=0,G_{g}=0\right]  $.
These are population outcome regressions for the never-treated group and for
the \textquotedblleft not-yet-treated\textquotedblright\ by time $t+\delta$
group. Let
\begin{align}
ATT_{ipw}^{nev}\left(  g,t;\delta\right)   &  =\mathbb{E}\left[  \left(
\frac{G_{g}}{\mathbb{E}\left[  G_{g}\right]  }-\frac{\dfrac{p_{g}\left(
X\right)  C}{1-p_{g}\left(  X\right)  }}{\mathbb{E}\left[  \dfrac{p_{g}\left(
X\right)  C}{1-p_{g}\left(  X\right)  }\right]  }\right)  \left(
Y_{t}-Y_{g-\delta-1}\right)  \right]  ,\label{eqn:att_ipw_nev}\\
ATT_{or}^{nev}\left(  g,t;\delta\right)   &  =\mathbb{E}\left[  \frac{G_{g}%
}{\mathbb{E}\left[  G_{g}\right]  }\left(  Y_{t}-Y_{g-\delta-1}-m_{g,t,\delta
}^{nev}\left(  X\right)  \right)  \right]  ,\label{eqn:att_or_nev}\\
ATT_{dr}^{nev}\left(  g,t;\delta\right)   &  =\mathbb{E}\left[  \left(
\frac{G_{g}}{\mathbb{E}\left[  G_{g}\right]  }-\frac{\dfrac{p_{g}\left(
X\right)  C}{1-p_{g}\left(  X\right)  }}{\mathbb{E}\left[  \dfrac{p_{g}\left(
X\right)  C}{1-p_{g}\left(  X\right)  }\right]  }\right)  \left(
Y_{t}-Y_{g-\delta-1}-m_{g,t,\delta}^{nev}\left(  X\right)  \right)  \right]  .
\label{eqn:att_dr_nev}%
\end{align}
Analogously, let%
\begin{align}
ATT_{ipw}^{ny}\left(  g,t;\delta\right)   &  =\mathbb{E}\left[  \left(
\frac{G_{g}}{\mathbb{E}\left[  G_{g}\right]  }-\frac{\dfrac{p_{g,t+\delta
}\left(  X\right)  \left(  1-D_{t+\delta}\right)  \left(  1-G_{g}\right)
}{1-p_{g,t+\delta}\left(  X\right)  }}{\mathbb{E}\left[  \dfrac{p_{g,t+\delta
}\left(  X\right)  \left(  1-D_{t+\delta}\right)  \left(  1-G_{g}\right)
}{1-p_{g,t+\delta}\left(  X\right)  }\right]  }\right)  \left(  Y_{t}%
-Y_{g-\delta-1}\right)  \right]  ,\label{eqn:att_ipw_ny}\\
ATT_{or}^{ny}\left(  g,t;\delta\right)   &  =\mathbb{E}\left[  \frac{G_{g}%
}{\mathbb{E}\left[  G_{g}\right]  }\left(  Y_{t}-Y_{g-\delta-1}-m_{g,t,\delta
}^{ny}\left(  X\right)  \right)  \right]  ,\label{eqn:att_or_ny}\\
ATT_{dr}^{ny}\left(  g,t;\delta\right)   &  =\mathbb{E}\left[  \left(
\frac{G_{g}}{\mathbb{E}\left[  G_{g}\right]  }-\frac{\dfrac{p_{g,t+\delta
}\left(  X\right)  \left(  1-D_{t+\delta}\right)  \left(  1-G_{g}\right)
}{1-p_{g,t+\delta}\left(  X\right)  }}{\mathbb{E}\left[  \dfrac{p_{g,t+\delta
}\left(  X\right)  \left(  1-D_{t+\delta}\right)  \left(  1-G_{g}\right)
}{1-p_{g,t+\delta}\left(  X\right)  }\right]  }\right)  \left(  Y_{t}%
-Y_{g-\delta-1}-m_{g,t,\delta}^{ny}\left(  X\right)  \right)  \right]  .
\label{eqn:att_dr_ny}%
\end{align}

With some abuse of notation,  we write $\bar{g} - \delta = \infty$ for any non-negative $\delta$ whenever $\bar{g} = \infty$.
\begin{theorem}
\label{thm:att} Let Assumptions \ref{ass:irrev}, \ref{ass:iid},
\ref{ass:anticipation} and \ref{ass:overlap} hold.

$\left(  i\right)  $ If Assumption \ref{ass:common-trends} holds, then, for
all $g$ and $t$ such that $g\in\mathcal{G}_{\delta}$, $t\in\left\{
2,\dots\mathcal{T-\delta}\right\}  $ and $ t \ge g-\delta$,%
\[
ATT\left(  g,t\right)  =ATT_{ipw}^{nev}\left(  g,t;\delta\right)
=ATT_{or}^{nev}\left(  g,t;\delta\right)  =ATT_{dr}^{nev}\left(
g,t;\delta\right)  .
\]

$\left(  ii\right)  $ If Assumption \ref{ass:common-trends-ny} holds, then,
for all $g$ and $t$ such that $g\in\mathcal{G}_{\delta}$, $t\in\left\{
2,\dots\mathcal{T-\delta}\right\}  $ and $ g-\delta \le t <\bar{g}-\delta$,
\[
ATT\left(  g,t\right)  =ATT_{ipw}^{ny}\left(  g,t;\delta\right)
=ATT_{or}^{ny}\left(  g,t;\delta\right)  =ATT_{dr}^{ny}\left(  g,t;\delta
\right)  .
\]

\end{theorem}

Theorem \ref{thm:att} is the first main result of this paper. It provides
powerful identification results that extend the DiD identification results
based on the outcome regression approach of \cite{Heckman1997, Heckman1998},
the IPW approach of \cite{Abadie2005}, and the DR approach of
\cite{SantAnna2020} to the multiple-periods, multiple groups setup. In other
words, Theorem \ref{thm:att} says that, from an identification point of view,
one can recover the $ATT\left(  g,t\right)  $'s by exploiting different parts
of the data generating process: the OR approach only relies on modeling the
conditional expectation of the outcome evolution for the comparison groups,
the IPW approach relies on modeling the conditional probability of being in group $g$,
whereas the DR approach exploits both OR and IPW components.

In order to extend the results of \cite{Heckman1997, Heckman1998},
\cite{Abadie2005}, and \cite{SantAnna2020} to the multiple groups, multiple
periods framework, we have to address two different challenges: one associated
with an appropriate reference time period and one associated with an
appropriate comparison group. Theorem \ref{thm:att} highlights how a solution
to these challenges is directly connected to the limited anticipation and the
conditional parallel trends assumptions. More specifically, Theorem
\ref{thm:att} says that we can use the time period $t=g-\delta-1$ as an
appropriate reference time period under Assumption \ref{ass:anticipation} and
either Assumption \ref{ass:common-trends} or \ref{ass:common-trends-ny}. This
is the most recent time period when untreated potential outcomes are observed
for units in group $g$. Interestingly, the more treatment anticipation is
allowed (i.e., the higher $\delta$ is), the further back in time one needs to
go.\footnote{As mentioned in Remark \ref{Rem:anticip}, as one allows $\delta$
to increase, Assumptions \ref{ass:common-trends} and
\ref{ass:common-trends-ny} becomes more restrictive.} Theorem \ref{thm:att}
also suggests that the choice of comparison group is directly tied to the
conditional parallel trends assumption one makes: under Assumption
\ref{ass:common-trends}, one can use \textquotedblleft never
treated\textquotedblright\ units as a fixed comparison group for all
\textquotedblleft eventually treated\textquotedblright\ units; whereas, under
Assumption \ref{ass:common-trends-ny}, one can use the \textquotedblleft
not-yet-treated by time $t+\delta$\textquotedblright\ units as a valid
comparison group for those who are first treated at time $g$. In this latter case, Theorem \ref{thm:att} also highlights that when all units eventually gets treated ($\bar{g} <\infty$), one is only able to identify the $ATT(g,t)$'s for time periods before the last treated group ``effectively'' starts their treatment, i.e., $t < \bar{g} - \delta$. In this case, one can not identify the $ATT(g,t)$ for the last treated cohort, too.

Finally, we note that when pre-treatment covariates play no role in
identification (i.e., Assumptions \ref{ass:anticipation},
\ref{ass:common-trends}, and \ref{ass:common-trends-ny} hold unconditionally on
$X$), (\ref{eqn:att_ipw_nev})-(\ref{eqn:att_dr_nev}) collapse to%
\begin{equation}
ATT_{unc}^{nev}(g,t;\delta)=E[Y_{t}-Y_{g-\delta-1}|G_{g}=1]-E[Y_{t}%
-Y_{g-\delta-1}|C=1], \label{eqn:attgt-nocovs}%
\end{equation}
and (\ref{eqn:att_ipw_ny})-(\ref{eqn:att_dr_ny}) collapse to
\begin{equation}
ATT_{unc}^{ny}(g,t;\delta)=E[Y_{t}-Y_{g-\delta-1}|G_{g}=1]-E[Y_{t}%
-Y_{g-\delta-1}|D_{t+\delta}=0]. \label{eqn:attgt-nocovs-ny}%
\end{equation}
These expressions for $ATT(g,t)$ clearly resemble the one for $ATT$ in the
canonical two-periods and two-groups case. As in that case, the average effect
of participating in the treatment for units in group $g$ is identified by
taking the path of outcomes (i.e., the change in outcomes between the most
recent period before they were affected by the treatment and the current
period) actually experienced by that group and adjusting it by the path of
outcomes experienced by a comparison group. Under the parallel trends
assumption, this latter path is the path of outcomes that units in group $g$
would have experienced if they had not participated in the treatment.

\begin{remark}
\label{Rem:unc}From (\ref{eqn:attgt-nocovs}) one can see that when Assumptions
\ref{ass:anticipation} and \ref{ass:common-trends} hold unconditionally and
there is no-anticipation, the $ATT\left(  g,t\right)  $ parameter can be
obtained by first subsetting the data to only contain observations at time $t$
and $g-1$, from units with either $G_{g}=1$ or $C=1$, and then, using only
the observations of this subset, running the (population) linear regression%
\begin{equation}
Y=\alpha_{1}^{g,t}+\alpha_{2}^{g,t}\cdot G_{g}+\alpha_{3}^{g,t}\cdot1\left\{
T=t\right\}  +\beta^{g,t}\cdot\left(  G_{g}\times1\left\{  T=t\right\}
\right)  +\epsilon^{g,t}. \label{eqn:regression}%
\end{equation}
It is then easy to verify that $\beta^{g,t}=ATT\left(  g,t\right)  $. Note
that one would need to consider different partitions of the data to
characterize different $ATT\left(  g,t\right)  $ in terms of regression
parameters. Alternatively, one could use the interacted two-way fixed effects
regression proposed by \cite{Abraham2018}.
\end{remark}

\begin{remark}
When covariates are available, the $\tilde{\beta}^{g,t}$ coefficient of the
population linear regression%
\[
Y=\tilde{\alpha}_{1}^{g,t}+\tilde{\alpha}_{2}^{g,t}\cdot G_{g}+\tilde{\alpha
}_{3}^{g,t}\cdot1\left\{  T=t\right\}  +\tilde{\beta}^{g,t}\cdot\left(
G_{g}\times1\left\{  T=t\right\}  \right)  +\tilde{\gamma}\cdot X+\tilde
{\epsilon}^{g,t}%
\]
that uses the same subset of data as in Remark \ref{Rem:unc} is, in general,
not equal to $ATT\left(  g,t\right)  $ unless one is willing to assume
$\left(  i\right)  $ homogeneous (in $X$) treatment effects, i.e.,
$\mathbb{E}[Y_{t}(g)-Y_{t}(0)|G_{g}=1,X]=\mathbb{E}[Y_{t}(g)-Y_{t}%
(0)|G_{g}=1]$ a.s., and $\left(  ii\right)  $ rule-out covariate-specific
trends, i.e., for $\mathbb{E}\left[  Y_{t}-Y_{t-1}|X,G\right]  =\mathbb{E}%
[Y_{t}-Y_{t-1}|G]$ a.s. for all groups and time periods; see, e.g.,
\cite{sloczynski-2018} for a related discussion. The characterizations of
$ATT\left(  g,t\right)  $ discussed in Theorem \ref{thm:att} do not rely on
these restrictive conditions.
\end{remark}

\begin{remark}
Although the IPW, OR, and DR based estimands presented in Theorem \ref{thm:att}
are identical from an identification standpoint, this is not the case when one
wants to estimate and make inference about the $ATT\left(  g,t\right)  $. As
we discuss in Section \ref{sec:estimation}, DiD  estimators based on the DR
estimands (\ref{eqn:att_dr_nev}) and (\ref{eqn:att_dr_ny}) usually enjoy
additional robustness against model-misspecifications when compared to the IPW
and OR estimands.
\end{remark}

\begin{remark}
Theorem \ref{thm:att} suggests that we can identify $ATT(g,t)$ only for groups in $\mathcal{G}_\delta \subseteq \mathcal{G}$ which can involve dropping some ``early treated'' groups due to anticipation effects.  When $\delta=0$, i.e.\ when there is no anticipation, $\mathcal{G}_\delta = \mathcal{G}$.  Theorem \ref{thm:att} also suggests that we can identify $ATT\left(  g,t\right)  $
only until $t=\mathcal{T}-\delta$ because of potential treatment anticipation
behavior. In applications {where some units are known to never participate in
the treatment (including periods after time period $\mathcal{T}$),} however,
we can identify $ATT(g,t)$ up to $t=\mathcal{T}$ by using these units as a
valid comparison group for all time periods $t=\mathcal{T}-\delta
+1,\dots,\mathcal{T}$, provided that an appropriate parallel trends assumption
is satisfied.
\end{remark}

\begin{remark}
From Theorem \ref{thm:att} it is clear that pre-treatment covariates play a
prominent role in our analysis. Importantly, Assumptions
\ref{ass:common-trends} and \ref{ass:common-trends-ny} suggest that
researchers should include pre-treatment covariates that are
potentially associated with the outcome evolution of $Y\left(  0\right)  $
during post-treatment periods. We explicitly rule out incorporating
post-treatment covariates as they can potentially be affected by the
treatment; see, e.g., \cite{Wooldridge2005a}, for a related discussion under
the unconfoundedness setup.
\end{remark}

\section{Summarizing Group-Time Average Treatment Effects {\label{sec:4} }}

The previous section shows that we can identify the $ATT(g,t)$'s by
restricting treatment anticipation behavior and imposing a conditional
parallel trends assumption. In many applications, the $ATT(g,t)$'s can be the
ultimate causal parameters of interest.  They can be used to highlight
treatment effect heterogeneity across different groups $g$, at different
points in time $t$, and across different lengths of treatment exposure,
$e=t-g$. In other situations, however, researchers may want to combine these
different $ATT\left(  g,t\right)  $'s to form more aggregated causal
parameters. For instance, if the number of groups and time periods is
relatively large, it may be challenging to interpret many group-time average
treatment effects.

In this section, we consider different aggregation schemes for the
$ATT(g,t)$'s that allow researchers to form a variety of summary measures of
the causal effects of a given policy. Our
aggregation schemes are of the form
\begin{equation}
	\theta=\sum_{g\in\mathcal{G}}\sum_{t=2}^{\mathcal{T}}w\left(  g,t\right)
	\cdot ATT(g,t), \label{eq:aggte}%
\end{equation}
where $w\left(  g,t\right)  $ are carefully-chosen (known or estimable) weighting functions
specified by the researcher such that $\theta$ can be used to address a
well-posed empirical/policy question.  Difference choices of $w\left(  g,t\right)$ allows researchers to highlight different types of treatment effect heterogeneity. We pay particular attention to aggregations that result in a single overall treatment effect summary parameter as well as to aggregations related to understanding dynamic effects as is commonly done in event-study analysis. Of course, many other aggregated parameters of the type (\ref{eq:aggte}) can be easily constructed following our framework. We illustrate this point by also summarizing heterogeneity with respect to group or by calendar time.

Before proceeding with the discussion on how to construct these different
aggregated parameters, it is worth revisiting the two most popular treatment
effect summary measures used by practitioners in DiD setups. These are based
on the \textquotedblleft static\textquotedblright\ and \textquotedblleft
dynamic\textquotedblright\ two-way fixed effects (TWFE) linear regression
specifications%
\begin{align}
Y_{i,t}  &  =\alpha_{t}+\alpha_{g}+\beta D_{i,t}+\epsilon_{i,t}%
,\label{eq:TWFE}\\
Y_{i,t}  &  =\alpha_{t}+\alpha_{g}+\sum_{e=-K}^{-2}\delta_{e}^{anticip}\cdot
D_{i,t}^{e}+\sum_{e=0}^{L}\beta_{e}\cdot D_{i,t}^{e}+v_{i,t},
\label{eq:TWFE_dyn}%
\end{align}
respectively, where $\alpha_{t}$ is a time fixed effect, $\alpha_{g}$ is a
group fixed effect, $\epsilon_{i,t}$ and $v_{i,t}$ are error terms,
$D_{i,t}^{e}=1\left\{  t-G_{i}=e\right\}  $ is an an indicator for unit $i$
being $e$ periods away from initial treatment at time $t$, and $K$ and $L$ are
positive constants. The parameter of interest in the static TWFE specification
is $\beta$, which, in applications, is typically interpreted as an overall
effect of participating in the treatment across groups and time periods. In
the dynamic TWFE specification, practitioners usually focus on the $\beta_{e}%
$, $e\geq0,$ and these parameters are typically interpreted as measuring the effect of participating in the treatment at different lengths of exposure to the treatment.

Despite the popularity of these specifications, recent research has shown that
one must be very careful in attaching a causal interpretation to these
aggregated parameters. For instance, \cite{Borusyak2017},
\cite{Goodman-bacon2017}, \cite{DeChaisemartin2016}, and \cite{Athey2018} have
shown that, in general, $\beta$ recovers a weighted average of some underlying
treatment effect parameters but some of the weights on these parameters can be
negative. This can potentially lead to particularly problematic cases such as
the effect of the treatment being positive for all units, but the TWFE
estimation resulting in estimates of $\beta$ that are negative. Even in cases
where the weights are not negative, the weights on underlying treatment effect
parameters are entirely driven by the TWFE estimation strategy and are
sensitive to the size of each group, the timing of treatment, and the total
number of time periods (see Theorem 1 in \cite{Goodman-bacon2017}). The
results in this section can be used in exactly the same setup to identify a
single interpretable average treatment effect parameter and, thus, provide a
way to circumvent the issues with the more common approach.

As discussed by \cite{Goodman-bacon2017}, the \textquotedblleft negative
weight problem\textquotedblright\ associated with $\beta$ arises when
treatment effects evolve over time. Thus, one may wonder if such problems
would still be present when considering more general, dynamic specifications
such as (\ref{eq:TWFE_dyn}). \cite{Abraham2018} shows that this is still the
case as the $\beta_{e}^{\prime}s$ associated with (\ref{eq:TWFE_dyn}) do not
recover easy-to-interpret causal parameters and still generally suffer from
the same sorts of \textquotedblleft negative weighting
problems.\textquotedblright\ In contrast to this, we provide a simple way to
directly aggregate our group-time average treatment effects into average
treatment effects across different lengths of exposure to the treatment.

\subsection{Aggregations to Highlight Treatment Effect Heterogeneity}

Next, we discuss several partial aggregations of the group-time average
treatment effects in order to summarize different dimensions of treatment
effect heterogeneity. Although there are additional possibilities, we focus
our discussion below on how to answer three particular questions: $\left(
a\right)  $ How does the effect of participating in the treatment vary with
length of exposure to the treatment? $\left(  b\right)  $ Do groups that are
treated earlier have, on average, higher/lower average treatment effects than
groups that are treated later? $\left(  c\right)  $ What is the cumulative
average treatment effect of the policy across all groups until some particular
point in time? Throughout this section, to avoid notation clutter, we assume that units
do not anticipate treatment, i.e., we consider the case where Assumption
\ref{ass:anticipation} holds with $\delta=0$. We also assume that a ``never treated'' group is available.

\subsubsection{How do average treatment effects vary with length of exposure
to the treatment?}

One of the most popular questions that arises in DiD setups with multiple time
periods concerns treatment effect dynamics: How does the effect of
participating in the treatment vary with length of exposure to the treatment?
For instance, do average treatment effects increase/decrease with elapsed
treatment time? Indeed, answering this type of question is often the main
motivation for using the event study regression in (\ref{eq:TWFE_dyn}),
though, as we mentioned above, that sort of regression may not be suitable for
such a task. In this section, we propose an aggregation scheme that is
suitable to highlight treatment effect heterogeneity with respect to length of
exposure to the treatment that does not suffer from the drawbacks associated
with the event study regression in (\ref{eq:TWFE_dyn}).

Let $e$ denote event-time, i.e., $e=t-g$ denotes the time elapsed since
treatment was adopted. Recall that $G$ denotes the time period that a unit is
first treated. Thus, a way to aggregate the $ATT(g,t)$'s to highlight
treatment effect heterogeneity with respect to $e$ is%
\begin{equation}
\theta_{es}(e)=\sum_{g\in\mathcal{G}}\mathbf{1}\{g+e\leq\mathcal{T}%
\}P(G=g|G+e\leq\mathcal{T})ATT(g,g+e). \label{eqn:agg-es}%
\end{equation}
This is the average effect of participating in the treatment $e$ time periods
after the treatment was adopted across all groups that are ever observed to
have participated in the treatment for exactly $e$ time periods. Here, the
\textquotedblleft on impact\textquotedblright\ average effect of participating
in the treatment occurs for $e=0$. $\theta_{es}(e)$ is the natural target for
event study regressions that are common in applied work, though it completely
avoids the pitfalls associated with the dynamic TWFE specification in
(\ref{eq:TWFE_dyn}).\footnote{Many of the parameters in this section involve
expressions that have similar components as the one for $\theta_{es}(e)$ in (\ref{eqn:agg-es}), and it is worth mentioning a few extra details for this case that are common to the other expressions below. The
term involving the indicator function, $\mathbf{1}\{g+e\leq \mathcal{T}\}$, limits consideration to identified
group-time average treatment effects.  The summation over groups with group specific weights, in this case given by $P(G=g|G+e\leq \mathcal{T})$, calculates an average, weighted by group size, of $ATT(g,t)$'s that are involved in a particular aggregation.  In addition, it is straightforward to
show that $\theta_{es}(e)$ can be written in the form of $\theta$ in Equation
(\ref{eq:aggte}). Throughout this section, we have written each parameter of
interest in its most intuitive form. Weights for each parameter in this section
corresponding to the form of the weights in Equation (\ref{eq:aggte}) are
provided in Table \ref{tab:agg-weights}.}

{ \setlength{\tabcolsep}{12pt} \newcolumntype{.}{D{.}{.}{-1}}
\ctable[caption={Weights on $ATT(g,t)$ for Aggregated Parameters},label=tab:agg-weights, pos=t, mincapwidth=.95\textwidth]{rl}{\tnote[]{\textit{Notes}: This table provides expressions for the weights on each $ATT(g,t)$ (as in Equation \ref{eq:aggte}) for each parameter discussed in this section.  In all cases except for $\theta_c^{cumu}(\tilde{t})$, the weights are all non-negative and sum to one.  For $\theta_c^{cumu}(\tilde{t})$, the weights are all non-negative but sum up to $\tilde{t}-1$ (rather than one), but this is just a reflection of $\theta_{c}^{cumu}\left(  \tilde{t}\right)  $ being a cumulative treatment effect measure.}}{\FL
\multicolumn{1}{c}{Parameter}&\multicolumn{1}{c}{$w(g,t)$}\ML
$\begin{aligned}[t]
\theta_{es}(e) \\
\theta_{es}^{bal}(e,e') \\
\theta_{sel}(\tilde{g}) \\
\theta_c(\tilde{t}) \\
\theta_c^{cumu}(\tilde{t}) \\
\theta_W^O \\
\theta_{sel}^O
\end{aligned} $ &
$\begin{aligned}[t]
w_{e}^{es}\left(  g,t\right) &= \mathbf{1}\{g+e\leq\mathcal{T}\}\mathbf{1}\{t-g=e\}P(G=g|G+e\leq\mathcal{T}) \\
w_{e}^{es,bal}\left(  g,t\right)  &= \mathbf{1}\{g+e^{\prime}\leq\mathcal{T}\}\mathbf{1}\{t-g=e\}P(G=g|G+e'\leq\mathcal{T}) \\
w_{\tilde{g}}^{s}\left(g,t\right)  &= \left.  \mathbf{1}\{t \geq g\}\mathbf{1}\left\{g=\tilde{g}\right\}  \right/  \left(  \mathcal{T}-g+1\right) \\
w_{\tilde{t}}^{c}\left(  g,t\right)  &= \mathbf{1}\{t \geq g\}\mathbf{1}\left\{  t=\tilde{t}\right\}  P(G=g|G\leq t) \\
w_{\tilde{t}}^{c,cumu}\left(  g,t\right) &=\mathbf{1}\{t \geq g\}\mathbf{1}\left\{  t\leq\tilde{t}\right\}  P(G=g|G\leq t) \\
w_{W}^O\left(  g,t\right) &= \mathbf{1}\{t \geq g\} P(G=g|G\leq\mathcal{T})  / \textstyle \sum_{g\in\mathcal{G}}\sum_{t=2}^{\mathcal{T}}\mathbf{1}\{t \geq g\}P(G=g|G\leq\mathcal{T})  \\
w_{sel}^O\left(  g,t\right) &= \left.  \mathbf{1}\{t \geq g\} P(G=g|G \leq \mathcal{T}) \right/  \left(  \mathcal{T}-g+1\right)
\end{aligned}$ \LL
}}

In event study regressions, it is common to plot $\beta_{e}$ across different
values of $e$ and to interpret differences as being due to treatment effect
dynamics. Similarly, one can plot $\theta_{es}(e)$ across different $e$'s to
better understand treatment effect dynamics. When doing so, it is important to
be aware that these comparisons may include compositional changes that can
complicate the interpretation of these parameters (note that the same
complications arise for event study regressions as well). To see this, for
$0\leq e_{1}<e_{2}\leq\mathcal{T}-2$, consider the difference between
$\theta_{es}(e_{2})$ and $\theta_{es}(e_1)$ which is given by%
\begin{align}
\theta_{es}(e_{2})-\theta_{es}(e_{1})  &  =\sum_{g\in\mathcal{G}}%
\mathbf{1}\{g+e_{1}\leq\mathcal{T}\}P(G=g|G+e_{1}\leq\mathcal{T}%
)\underset{\text{dynamic effect for group $g$}}{\underbrace{\left(
ATT(g,g+e_{2})-ATT(g,g+e_{1})\right)  }}\label{eqn:dyn-diff}\\
&  \hspace{10pt}+\sum_{g\in\mathcal{G}}\mathbf{1}\{g+e_{2}\leq\mathcal{T}%
\}\underset{\text{differences in weights}}{\underbrace{\left(  P(G=g|G+e_{2}%
\leq\mathcal{T})-P(G=g|G+e_{1}\leq\mathcal{T})\right)  }}ATT(g,g+e_{2}%
)\nonumber\\
&  \hspace{10pt}-\sum_{g\in\mathcal{G}}\underset{\text{different composition
of groups}}{\underbrace{\mathbf{1}\{\mathcal{T}-e_{2}<g\leq\mathcal{T}%
-e_{1}\}}}P(G=g|G+e_{1}\leq\mathcal{T})ATT(g,g+e_{2}).\nonumber
\end{align}
From the above decomposition it becomes clear that comparing $\theta_{es}(e)$
at two different values of $e$ provides a weighted average of the dynamic
effect of participating in the treatment -- the first component on the
right-hand side of (\ref{eqn:dyn-diff}) -- plus two extra undesirable terms.
Both of these undesirable terms are due to different compositions of groups
at different event times.\footnote{The composition changes mentioned
here arise due to the staggered adoption of the treatment. For example, when
$\mathcal{T}=3$, groups 2 and 3 both show up in the expression for
$\theta_{es}(0)$, but only group 2 shows up in the expression for $\theta
_{es}(1)$.} The first term arises because the weights at each length of
exposure differ due to the changing composition of groups at each event time.
The second term comes directly from different compositions of groups at each
length of exposure. These two additional terms may prevent one from
interpreting the differences in $\theta_{es}(e)$ across different values of
$e$ as being actual dynamic effects of participating in the treatment unless
one is willing to impose that $ATT(g,g+e)$ does not vary with $g$ for any
$e\geq0$; i.e., that dynamic effects are common across groups.\footnote{If
$ATT(g,g+e)$ does not vary with $g$ for any $e\geq0$, it is straightforward to
show that the last two terms of (\ref{eqn:dyn-diff}) sum up to 0.} However,
this sort of homogeneity condition may be deemed too strong in many applications.

A simple alternative causal parameter that can be used to highlight treatment
effect dynamics with respect to length of exposure to the treatment and does
not suffer from the issue of compositional changes highlighted in
(\ref{eqn:dyn-diff}) arises from \textquotedblleft balancing\textquotedblright%
\ the groups with respect to event time, i.e., to only aggregate the
$ATT\left(  g,t\right)  $'s for a fixed set of groups that are exposed to the
treatment for at least some particular number of time periods and thereby circumvent
the issue of compositional changes across different values of $e$. In
particular, for some event time $e^{\prime}$ with $0\leq e\leq e^{\prime}%
\leq\mathcal{T}-2$, let%
\begin{equation}
\theta_{es}^{bal}(e;e^{\prime})=\sum_{g\in\mathcal{G}}\mathbf{1}\{g+e^{\prime
}\leq\mathcal{T}\}ATT(g,g+e)P(G=g|G+e^{\prime}\leq\mathcal{T}).
\label{eqn:agg-es-bal}%
\end{equation}
Notice that the definition of $\theta_{es}^{bal}(e;e^{\prime})$ is very
similar to $\theta_{es}(e)$ except that it calculates the average group-time
average treatment effect for units whose event time is equal to $e$
\textit{and who are observed to participate in the treatment for at least
$e^{\prime}$ periods}. In this case, since the composition of groups is the
same across all values of $e$, the additional terms in (\ref{eqn:dyn-diff}) do
not show up at all and differences in $\theta_{es}^{bal}(e;e^{\prime})$ across
different values of $e$ cannot be due to differences in the composition of
groups at different values of $e$. As an example, when one is interested in
analyzing the evolution of treatment effects up to 5 periods after treatment
was implemented, one can set $e^{\prime}=5$ and, this way, the same groups of
units will be used when computing $\theta_{es}^{bal}(0;5),\theta_{es}%
^{bal}(1;5),\dots,$ $\theta_{es}^{bal}(5;5)$.

The price one pays for \textquotedblleft balancing\textquotedblright\ the
groups with respect to event time is that fewer groups are used to compute
these event-study-type estimands, which can lead to less informative
inference. Thus, in practice, one should consider this \textquotedblleft
robustness\textquotedblright\ versus \textquotedblleft
efficiency\textquotedblright\ trade-off when choosing between $\theta
_{es}^{bal}$ and $\theta_{es}$.

\begin{remark}
We note that $\theta_{es}^{bal}(e;e^{\prime})$ closely resembles the empirical
practice of only reporting event-study-type coefficients for the event periods
that do not suffer from compositional changes, see, e.g., \cite{McCrary2007}
and \cite{Bailey2015}. An important caveat is that our proposed
event-study-type estimands $\theta_{es}^{bal}(e;e^{\prime})$ are not based on
dynamic TWFE specifications akin to (\ref{eq:TWFE_dyn}), and therefore bypass
the pitfalls associated with (\ref{eq:TWFE_dyn}) highlighted by
\cite{Abraham2018}.
\end{remark}

\subsubsection{How do average treatment effects vary across groups?}

It is also straightforward to aggregate our group-time average treatment
effects to understand heterogeneity in the effect of participating in the
treatment across groups. Although understanding this sort of heterogeneity is
relatively less common in applied work than trying to understand dynamic
effects as discussed above, there are still a number of cases in economics
where understanding this sort of heterogeneity may be of interest. For
example, work on the effect of graduating during a recession on labor market
outcomes (\citet{Oreopoulos2012}) or the effect of job displacement across the
business cycle (\citet{Farber2017}) are related to heterogeneous effects
across groups. More generally, these parameters are useful for understanding
if the effect of participating in the treatment was larger for groups that are
treated earlier relative to groups that are treated later. In addition, in the
next section, these parameters will be the building block for our main measure
of the overall effect of participating in the treatment. To consider
heterogeneous effects across groups, we consider the following parameter
\begin{equation}
\theta_{sel}(\tilde{g})=\frac{1}{\mathcal{T}-\tilde{g}+1}\sum_{t=\tilde{g}%
}^{\mathcal{T}}ATT(\tilde{g},t). \label{eqn:agg-sel}%
\end{equation}
$\theta_{sel}(\tilde{g})$ is the average effect of participating in the
treatment among units in group $\tilde{g}$, across all their post-treatment periods.

\subsubsection{What is the cumulative average treatment effect of the policy
across all groups until time $\tilde{t}$?}

In some applications, researchers may want to construct an aggregated target
parameter to highlight treatment effect heterogeneity with respect to calendar
time. In economics, for example, researchers might wish to study heterogeneous
treatment effects across the business cycle. The
average effect of participating in the treatment in time period $t$ (across
groups that have adopted the treatment by period $t$) is given by
\begin{equation}
\theta_{c}(\tilde{t})=\sum_{g\in\mathcal{G}}\mathbf{1}\{\tilde{t}\geq
g\}P(G=g|G\leq\tilde{t})ATT(g,t) \label{eqn:agg-cal}%
\end{equation}
An extension to this parameter is to think about the \textit{cumulative}
effect of participating in the treatment up to some particular time period.
For instance, in active labor market applications, policy makers may want to
know the cumulative average effect of a given training program on earnings
from the year that the first group of people were trained until year
$\tilde{t}$. This would provide a measure of the cumulative earnings gains
induced by the training program. Alternatively, in health applications,
researchers may want to measure how many COVID-19 cases have been averted by
shelter-in-place orders up to day $\tilde{t}$. To consider the cumulative
effect, consider the following parameter
\begin{equation}
\theta_{c}^{cumu}\left(  \tilde{t}\right)  =\sum_{t=2}^{\tilde{t}}\theta
_{c}(t). \label{eqn:agg-cumu}%
\end{equation}
$\theta_{c}^{cumu}\left(  \tilde{t}\right)  $ can be interpreted as the
cumulative average treatment effect among the units that have been treated by
time $\tilde{t}$.

\subsection{Aggregations into Overall Treatment Effect
Parameters\label{sec:overall}}

Finally in this section, we consider some ideas for aggregating group time
average treatment effects into an overall effect of participating in the
treatment. One very simple idea is to just average all of the identified
group-time average treatment effects together; i.e., to consider the
parameter
\begin{align}
\label{eqn:avg-overall}\theta_{W}^{O}=\frac{1}{\kappa}\sum_{g\in\mathcal{G}%
}\sum_{t=2}^{\mathcal{T}}\mathbf{1}\{t\geq g\}ATT(g,t)P(G=g|G\leq\mathcal{T})
\end{align}
where $\kappa=\sum_{g\in\mathcal{G}}\sum_{t=2}^{\mathcal{T}}1\{t\geq
g\}P(G=g|G\leq\mathcal{T})$ (which ensures that the weights on $ATT(g,t)$ in
the second term sum up to one). $\theta_{W}^{O}$ is a weighted average of each
$ATT(g,t)$ putting more weight on $ATT(g,t)$'s with larger group sizes. Unlike
$\beta$ in the TWFE regression specification (\ref{eq:TWFE}), this simple
combination of $ATT(g,t)$'s immediately rules out troubling issues due to
negative weights; as a particular example, when the effect of participating in
the treatment is positive for all units, this aggregated parameter cannot be negative.

That being said, just requiring positive weights is a very minimal requirement
of a reasonable overall treatment effect parameter. For example, one drawback
of $\theta_{W}^{O}$ is that it systematically puts more weight on groups that
participate in the treatment for longer. Instead, we suggest the following
parameter as a general-purpose summary of the average effect of participating
in the treatment
\begin{align}
\label{eqn:sel-overall}\theta_{sel}^{O}=\sum_{g\in\mathcal{G}}\theta
_{sel}(g)P(G=g|G\leq\mathcal{T})
\end{align}
where $\theta_{sel}(g)$ is the average effect of participating in the
treatment for units in group $g$ as defined in Equation (\ref{eqn:agg-sel})
above. $\theta^{O}_{sel}$ first computes the average effect for each group
(across all time periods) and then averages these effects together across
groups to summarize the overall average effect of participating in the
treatment. Thus, $\theta^{O}_{sel}$ is the average effect of participating in
the treatment experienced by all units that ever participated in the
treatment. In this respect, its interpretation is the same as the ATT in the
canonical DiD setup with two periods and two groups.
This is an attractive property for a summary measure of the overall effect of
participating in the treatment in the context of multiple time periods and
variation in treatment timing.

Working by analogy, one can also define overall treatment effect parameters by
averaging $\theta_{es}(e)$ across all event times or $\theta_{c}(t)$ across
all time periods, i.e.,
\begin{align}
\label{eqn:time-overall}\theta^{O}_{es} = \frac{1}{\mathcal{T}-1}\sum
_{e=0}^{\mathcal{T}-2} \theta_{es}(e) \quad\quad\quad\quad\theta^{O}_{c} =
\frac{1}{\mathcal{T}-1}\sum_{t=2}^{\mathcal{T}} \theta_{c}(t)
\end{align}
In our view, the appeal of these aggregations is likely to be somewhat more
limited than that of $\theta^{O}_{sel}$ in most applications. For example, the
interpretation of $\theta^{O}_{es}$ is complicated by the issue of the changing
composition of groups across different values of $e$ discussed above (similar
arguments apply to $\theta^{O}_{c}$ as well).
As before, one can circumvent the issue of the changing composition of groups by
balancing the sample with respect to event time. A (local) single summary
parameter is given by
\begin{align}
\label{eqn:bal-overall}\theta^{O,bal}_{es}(e^{\prime}) = \frac{1}{e^{\prime
}+1} \sum_{e=0}^{e^{\prime}} \theta^{bal}_{es}(e,e^{\prime})
\end{align}
This is the average effect of participating in the treatment over the first
$e^{\prime}$ periods of exposure to the treatment. This is also a reasonable
alternative overall treatment effect parameter, but it should also be noted
that it is local to groups that participated in the treatment for at least
$e^{\prime}$ periods.

As a final comment, in general, none of the overall effect parameters
considered in this section are equal to each other except in the special case
where $ATT(g,t)$ is the same for all groups and all time periods. In that
case, all of the aggregated parameters, including $\beta$ from the TWFE
regression, are equal to each other.

\section{Estimation and Inference\label{sec:estimation}}

So far we have focused on the identification and aggregation stages of the
analysis. In this section, we show how one can build on these results to form
estimators for and conduct inference about the group-time average treatment effects
and their summary measures described in Section \ref{sec:4}. Given that
the $ATT\left(  g,t\right)  $'s are the main building blocks of our analysis,
we start with them. 

First, it is important to notice that our identification results in Theorem
\ref{thm:att} are constructive and suggest a simple and intuitive two-step
estimation strategy to estimate the $ATT\left(  g,t\right)  $'s. In the first
step, one estimates the nuisance functions for each group $g$ and time period
$t$ --- $p_{g}(x)$ and/or $m_{g,t,\delta}^{nev}\left(  X\right)  $ if one
relies on Assumption \ref{ass:common-trends}, and $p_{g,t+\delta}(x)$ and/or
$m_{g,t,\delta}^{ny}\left(  X\right)  $ if one relies on Assumption
\ref{ass:common-trends-ny}. In the second step, one plugs the fitted values of
these estimated nuisance functions into the sample analogue of the considered
$ATT\left(  g,t\right)  $ estimand to obtain estimates of the group-time
average treatment effect.

A natural question that then arises is which type of approach one should use
in practice: the outcome regression, inverse probability weighting, or the
doubly-robust one. Although these three different approaches are equivalent
from the \emph{identification} perspective, this is not the case from the
\emph{estimation/inference} perspective. The OR approach requires researchers
to \emph{correctly} model the outcome evolution of the comparison group to
estimate the group-time average treatment effects. This approach is explicitly
connected with the conditional parallel trends assumption required in DiD 
analysis as this condition is usually expressed in terms of conditional
expectations. The IPW approach, on the other hand, avoids explicitly modeling
the outcome evolution of the comparison group and therefore does not rely on
putative model restrictions directly tied to the parameter of interest.
Instead, the IPW approach requires one to \emph{correctly }model the
conditional probability of unit $i$ being in group $g$ given their covariates
$X$ and that they are either in group $g$ or in an appropriate comparison
group. The DR approach combines both the OR and IPW approaches as it relies on
modeling both the outcome evolution and the propensity score. However, it only
requires one to correctly specify \emph{either (but not necessarily both) }the
outcome evolution for the comparison group or the propensity score model
\citep{SantAnna2020}. Thus, the DR approach enjoys additional robustness
against model misspecifications when compared to the OR and IPW approaches. In
addition, the DR approach potentially allows one to use a broader set of
estimation methods such as those that involve penalization and some types of
model selection, see, e.g. \cite{Belloni2017}.

Given these attractive robustness features associated with the DR approach, in
this section we consider estimators of the DR form; the discussion on how to
proceed with the OR and IPW approaches is analogous and therefore omitted. We
also focus on parametric estimators for the nuisance functions.  We consider this case mainly for its practical appeal which is especially true in applications where the number of covariates is fairly large and the number of observations is only moderate.\footnote{Alternatively, one could adopt a fully nonparametric approach. Let $f\left( x\right) $ be a generic notation for the nuisance functions. From \cite{Newey1994}, \cite{Chen2003}, \cite{Ai2003, Ai2007,
		Ai2012}, and \cite{Chen2008}, one can see that the use of nonparametric
	first-step estimators $\widehat{g}\left( x\right) $ of $g\left( x\right) $
	is warranted provided that $\left\Vert \widehat{g}\left( x\right) -g\left(
	x\right) \right\Vert _{\mathcal{H}}=o_{p}\left( n^{-1/4}\right) $ for a
	pseudo-metric $\left\Vert \cdot \right\Vert _{\mathcal{H}}$, $\mathcal{H}$
	being a vector space of functions. However, when the dimension of $X$ is
	moderate or large, as is often the case in empirical applications,
	conditions ensuring that $\left\Vert \widehat{g}\left( x\right) -g\left(
	x\right) \right\Vert _{\mathcal{H}}=o_{p}\left( n^{-1/4}\right) $ can be
	rather stringent due to the so-called ``curse of dimensionality''.}

More concisely, let%

\begin{align}
\widehat{ATT}_{dr}^{nev}\left(  g,t;\delta\right)   &  =\mathbb{E}_{n}\left[
\left(  \widehat{w}_{g}^{treat}-\widehat{w}_{g}^{comp,nev}\right)  \left(
Y_{t}-Y_{g-\delta-1}-\widehat{m}_{g,t,\delta}^{nev}\left(  X;\widehat{\beta
}_{g,t,\delta}^{nev}\right)  \right)  \right]  ,\label{eqn:ATT_dr_never}\\
\widehat{ATT}_{dr}^{ny}\left(  g,t;\delta\right)   &  =\mathbb{E}_{n}\left[
\left(  \widehat{w}_{g}^{treat}-\widehat{w}_{g}^{comp,ny}\right)  \left(
Y_{t}-Y_{g-\delta-1}-\widehat{m}_{g,t,\delta}^{ny}\left(  X;\widehat{\beta
}_{g,t,\delta}^{ny}\right)  \right)  \right]  , \label{eqn:ATT_dr_ny}%
\end{align}
where 
\small\[
\widehat{w}_{g}^{treat}=\frac{G_{g}}{\mathbb{E}_{n}\left[  G_{g}\right]
},~\widehat{w}_{g}^{comp,nev}=\frac{\dfrac{\widehat{p}_{g}\left(
X;\widehat{\pi}_{g}\right)  C}{1-\widehat{p}_{g}\left(  X;\widehat{\pi}%
_{g}\right)  }}{\mathbb{E}_{n}\left[  \dfrac{\widehat{p}_{g}\left(
X;\widehat{\pi}_{g}\right)  C}{1-\widehat{p}_{g}\left(  X;\widehat{\pi}%
_{g}\right)  }\right]  },~\widehat{w}_{g}^{comp,ny}=\frac{\dfrac
{\widehat{p}_{g,t+\delta}\left(  X;\widehat{\pi}_{g,t+\delta}\right)  \left(
1-D_{t+\delta}\right)  \left(  1-G_{g}\right)  }{1-\widehat{p}_{g,t+\delta
}\left(  X;\widehat{\pi}_{g,t+\delta}\right)  }}{\mathbb{E}_{n}\left[
\dfrac{\widehat{p}_{g,t+\delta}\left(  X;\widehat{\pi}_{g,t+\delta}\right)
\left(  1-D_{t+\delta}\right)  \left(  1-G_{g}\right)  }{1-\widehat{p}%
_{g,t+\delta}\left(  X;\widehat{\pi}_{g,t+\delta}\right)  }\right]  },
\]
\normalsize
with $\widehat{p}_{g}\left(  \cdot;\widehat{\pi}_{g}\right)  $,
$\widehat{p}_{g,t+\delta}(\cdot;\widehat{\pi}_{g,t+\delta})$, $\widehat{m}%
_{g,t,\delta}^{nev}(\cdot;\widehat{\beta}_{g,t,\delta}^{nev})$ and
$\widehat{m}_{g,t,\delta}^{ny}(\cdot;\widehat{\beta}_{g,t,\delta}^{ny})$ being
(parametric) estimators of $p_{g}(\cdot)$, $p_{g,t+\delta}(\cdot)$,
$m_{g,t,\delta}^{nev}(\cdot)$ and $m_{g,t,\delta}^{ny}(\cdot)$, respectively,
and for a generic $Z$, $\mathbb{E}_{n}\left[  Z\right]  =n^{-1}\sum_{i=1}%
^{n}Z_{i}$. $\widehat{ATT}_{dr}^{nev}\left(  g,t;\delta\right)  $ and
$\widehat{ATT}_{dr}^{nev}\left(  g,t;\delta\right)  $ are our proposed DR DiD 
estimators for $ATT\left(  g,t\right)  $ when one invokes Assumption
\ref{ass:common-trends} and Assumption \ref{ass:common-trends-ny},
respectively. These estimators extend the DR DiD estimators of
\cite{SantAnna2020} from the two periods, two groups setup to the multiple
groups, multiple periods setup while allowing for possible treatment
anticipation. In addition, these estimators are of the \cite{Hajek1971}-type
and their associated weights are guaranteed to sum up to one in finite
samples. As illustrated by \cite{Busso2014}, this usually leads to improved
finite sample properties.

With the estimators for the $ATT\left(  g,t\right)  $'s in hand, one can use
the analogy principle and combine these to estimate the summarized average
treatment effect parameters discussed in Section \ref{sec:4}.

\begin{remark}
In applications with limited covariate overlap (i.e., with propensity scores
sufficiently close to one), IPW and DR estimators may lead to imprecise
(irregular) inference procedures, see, e.g., \cite{Khan2010}. In such cases,
provided that one is comfortable with (parametric) extrapolation and is
sufficiently confident that the outcome regression working models are
correctly specified, relying on the OR estimation approach may lead to more
informative inferences. Alternatively, one may choose to trim extreme
propensity score estimates though proceeding in this manner would change the
target parameter; i.e., we would not be recovering the $ATT\left(  g,t\right)
$'s; see, e.g., \cite{Crump2009} and \cite{Yang2018} for related discussion in
other contexts. In the rest of the paper, we abstract from these points.
\end{remark}

\subsection{Asymptotic Theory for Group-Time Average Treatment
Effects\label{sec:asy-agg}}

Next, we derive the asymptotic properties of our DR DiD estimators for the
$ATT\left(  g,t\right)  $'s. To simplify exposition, we focus on the case with
a never-treated comparison group as in (\ref{eqn:ATT_dr_never}); results that
come from using the not-yet-treated group as the comparison group as in
(\ref{eqn:ATT_dr_ny}) follow from symmetric arguments and are therefore omitted. We also note that the theoretical results in this section are justified within the large $n$, fixed $\mathcal{T}$ paradigm.

Let $\left\Vert Z\right\Vert =\sqrt{trace\left(  Z^{\prime}Z\right)  }$ denote
the Euclidean norm of $Z$ and set $W=\left(  Y_{1},\dots,Y_{\mathcal{T}%
},X,D_{1},\dots,D_{\mathcal{T}}\right)  $. For a generic $\kappa_{g,t}^{nev}=\left(
\pi_{g}^{\prime},\beta_{g,t,\delta}^{nev}\right)  ^{\prime}$, let%
\[
h_{g,t}^{dr,nev}\left(  W;\kappa_{g,t}^{nev},\delta\right)  =\left(
w_{g}^{treat}\left(  W\right)  -w_{g}^{comp,nev}\left(  W;\pi_{g}\right)
\right)  \left(  Y_{t}-Y_{g-\delta-1}-m_{g,t,\delta}^{nev}\left(
X;\beta_{g,t,\delta}^{nev}\right)  \right)  ,
\]
where the normalized weights $w_{g}^{treat}\left(  W\right)  $ and
$w_{g}^{comp,nev}\left(  W;\pi_{g}\right)  $ are given by%
\begin{equation}
w_{g}^{treat}\left(  W\right)  =\frac{G_{g}}{\mathbb{E}\left[  G_{g}\right]
},~~~~~~~~~~~~w_{g}^{comp,nev}\left(  W;\pi_{g}\right)  =\left.  \dfrac
{p_{g}\left(  X;\pi_{g}\right)  C}{1-p_{g}\left(  X;\pi_{g}\right)  }\right/
\mathbb{E}\left[  \dfrac{p_{g}\left(  X;\pi_{g}\right)  C}{1-p_{g}\left(
X;\pi_{g}\right)  }\right]  . \label{eqn:weights}%
\end{equation}
Let $g\left(  \cdot\right)  $ be a generic notation for $p_{g}\left(
\cdot\right)  $ and $m_{g,t,\delta}^{nev}\left(  \cdot\right)  $. With some
abuse of notation, let $g\left(  \cdot;\gamma\right)  $ be a generic notation
for $p_{g}\left(  \cdot;\pi_{g}\right)  $ and $m_{g,t,\delta}^{nev}\left(
\cdot;\beta_{g,t,\delta}^{nev}\right)  $. The vector of pseudo-true parameters
is given by $\kappa_{g,t}^{\ast,nev}=\left(  \pi_{g}^{\ast\prime}%
,\beta_{g,t,\delta}^{\ast,nev~\prime}\right)  ^{\prime}$. Finally, let
$\dot{h}_{g,t}^{dr,nev}\left(  W;\kappa_{g,t}^{nev}\right)  =\partial\left.
h_{g,t}^{dr,nev}\left(  W;\kappa_{g,t}^{nev}\right)  \right/  \partial
\kappa_{g,t}^{nev}$.

\begin{assumption}
\label{ass:parametric} $\left(  i\right)  $ $g\left(  x;\gamma\right)  $ is a
parametric model for $g\left(  x\right)  $, where $\gamma\in\Theta
\subset\mathbb{R}^{k}$, $\Theta$ being compact; $\left(  ii\right)  $
$g\left(  X;\gamma\right)  $ is $a.s.$ continuous at each $\gamma$ $\in\Theta
$; $\left(  iii\right)  $ there exists a unique pseudo-true parameter
$\gamma^{\ast}\in int\left(  \Theta\right)  $; $\left(  iv\right)  $ $g\left(
X;\gamma\right)  $ is $a.s.$ twice continuously differentiable in a
neighborhood of $\gamma^{\ast}$, $\Theta^{\ast}\subset\Theta$; $\left(
v\right)  $ the estimator $\widehat{\gamma}$ is strongly consistent for
$\gamma^{\ast}$ and satisfies the following linear expansion:%
\[
\sqrt{n}\left(  \widehat{\gamma}-\gamma^{\ast}\right)  =\frac{1}{\sqrt{n}}%
\sum_{i=1}^{n}l_{g,t}\left(  W_{i};\gamma^{\ast}\right)  +o_{p}\left(
1\right)  ,
\]
where $l_{g,t}\left(  \cdot;\gamma\right)  $ is a $k\times1$ vector such that
$\mathbb{E}\left[  l_{g,t}\left(  W;\gamma^{\ast}\right)  \right]  =0$,
$\mathbb{E}\left[  l_{g,t}\left(  W;\gamma^{\ast}\right)  l_{g,t}\left(
W;\gamma^{\ast}\right)  ^{\prime}\right]  $ exists and is positive definite
and $\lim_{s\rightarrow0}\mathbb{E}\left[  \sup_{\gamma\in\Theta^{\ast
}:\left\Vert \gamma-\gamma^{\ast}\right\Vert \leq s}\left\Vert l_{g}\left(
W;\gamma\right)  -l_{g}\left(  W;\gamma^{\ast}\right)  \right\Vert
^{2}\right]  =0$. In addition, $\left(  vi\right)  $~for some $\varepsilon>0$
and all $g\in\mathcal{G}$, $0\leq p_{g}\left(  X;\pi_{g}\right)
\leq1-\varepsilon$ $a.s.$, for all $\pi\in int\left(  \Theta^{ps}\right)  $,
where $\Theta^{ps}$ denotes the parameter space of $\pi_{g}.$
\end{assumption}

\begin{assumption}
\label{ass:integrabilitity} For each $g\in\mathcal{G}$ and $t=\left\{
2,\dots,\mathcal{T-\delta}\right\}  $, assume that $\mathbb{E}\left[
\left\Vert h_{g,t}^{nev}\left(  W;\kappa_{g,t}^{\ast,nev},\delta\right)
\right\Vert ^{2}\right]  <\infty$ and $\mathbb{E}\left[  \sup_{\kappa\in
\Gamma^{\ast}}\left\vert \dot{h}_{g,t}^{nev}\left(  W;\kappa\right)
\right\vert \right]  <\infty,$ where $\Gamma^{\ast}$ is a small neighborhood
of $\kappa_{g,t}^{\ast,nev}$.
\end{assumption}

Assumptions \ref{ass:parametric}-\ref{ass:integrabilitity} are standard in the
literature, see e.g.~\cite{Abadie2005}, \cite{Wooldridge2007a},
\cite{Bonhomme2011}, \cite{Graham2012}, and \cite{SantAnna2020}. Assumption
\ref{ass:parametric} requires that the first-step estimators are based on
smooth parametric models and that the estimated parameters admit $\sqrt{n}%
$-asymptotically linear representations, whereas Assumption
\ref{ass:integrabilitity} imposes some weak integrability conditions. Under
mild moment conditions, these requirements are fulfilled when one adopts
linear/nonlinear outcome regressions or logit/probit models, for example, and
estimates the unknown parameters by (nonlinear) least squares, quasi-maximum
likelihood, or other alternative estimation methods, see e.g.~Chapter 5 in
\cite{VanderVaart1998}, \cite{Wooldridge2007a}, \cite{Graham2012} and
\cite{SantAnna2020}. In other words, Assumptions \ref{ass:parametric}%
-\ref{ass:integrabilitity} allow for flexible parametric specifications of the
nuisance functions and accommodate different estimation methods.

In what follows, we write $w_{g}^{treat}=w_{g}^{treat}\left(  W\right)  $,
$w_{g}^{comp}\left(  \pi_{g}\right)  =w_{g}^{comp,nev}\left(  W;\pi
_{g}\right)  $, and $m_{g,t,\delta}^{nev}\left(  \beta_{g,t,\delta}%
^{nev}\right)  =m_{g,t,\delta}^{nev}\left(  X;\beta_{g,t,\delta}^{nev}\right)
$ to minimize notation. For a generic $\kappa_{g,t}^{nev}=\left(  \pi
_{g}^{\prime},\beta_{g,t,\delta}^{nev^{\prime}}\right)  ^{\prime}$, define%
\begin{equation}
\psi_{g,t,\delta}^{dr,nev}(W_{i};\kappa_{g,t}^{nev})=\psi_{g,t,\delta
}^{treat,nev}(W_{i};\beta_{g,t,\delta}^{nev})-\psi_{g,t,\delta}^{comp,nev}%
(W_{i};\pi_{g},\beta_{g,t,\delta}^{nev})-\psi_{g,t,\delta}^{est,nev}(W_{i}%
;\pi_{g},\beta_{g,t,\delta}^{nev}), \label{eqn:psigt}%
\end{equation}
with
\begin{align*}
\psi_{g,t,\delta}^{treat,nev}(W;\beta_{g,t,\delta}^{nev})  &  =w_{g}%
^{treat}\cdot\left(  Y_{t}-Y_{g-\delta-1}-m_{g,t,\delta}^{nev}\left(
\beta_{g,t,\delta}^{nev}\right)  \right) \\
&  ~~~~~~~~-w_{g}^{treat}\cdot\mathbb{E}\left[  w_{g}^{treat}\cdot\left(
Y_{t}-Y_{g-\delta-1}-m_{g,t,\delta}^{nev}\left(  \beta_{g,t,\delta}%
^{nev}\right)  \right)  \right]  ,\\
\psi_{g,t,\delta}^{comp,nev}(W;\pi_{g},\beta_{g,t,\delta}^{nev})  &
=w_{g}^{comp}\left(  \pi_{g}\right)  \cdot\left(  Y_{t}-Y_{g-\delta
-1}-m_{g,t,\delta}^{nev}\left(  \beta_{g,t,\delta}^{nev}\right)  \right) \\
&  ~~~~~~~~-w_{g}^{comp}\left(  \pi_{g}\right)  \cdot\mathbb{E}\left[
w_{g}^{comp}\left(  \pi_{g}\right)  \cdot\left(  Y_{t}-Y_{g-\delta
-1}-m_{g,t,\delta}^{nev}\left(  \beta_{g,t,\delta}^{nev}\right)  \right)
\right]  ,
\end{align*}
and%
\[
\psi_{g,t}^{est,nev}(W;\pi_{g},\beta_{g,t,\delta}^{nev})=l_{g,t}%
^{or,nev}\left(  \beta_{g,t,\delta}^{nev}\right)  ^{\prime}\cdot
M_{g,t,\delta}^{dr,nev,1}+l_{g}^{ps,nev}\left(  \pi_{g}\right)  ^{\prime}\cdot
M_{g,t,\delta}^{dr,nev,2},
\]
where $l_{g,t}^{or,nev}\left(  \cdot\right)  $ is the asymptotic linear
representation of the estimator for the outcome evolution of the comparison
groups as described in Assumption \ref{ass:parametric}$\left(  iv\right)  $,
$l_{g}^{ps,nev}\left(  \cdot\right)  $ is defined analogously for the
generalized propensity score, and
\begin{align*}
M_{g,t,\delta}^{dr,nev,1}  &  =\mathbb{E}\left[  \left(  w_{g}^{treat}%
-w_{g}^{comp}\left(  \pi_{g}\right)  \right)  \cdot\dot{m}_{g,t,\delta}%
^{nev}\left(  \beta_{g,t,\delta}^{nev}\right)  \right]  ,\\
M_{g,t,\delta}^{dr,nev,2}  &  =\mathbb{E}\left[  \alpha_{g}^{ps,nev}\left(
\pi_{g}\right)  \cdot\left(  Y_{t}-Y_{g-\delta-1}-m_{g,t,\delta}^{nev}\left(
\beta_{g,t,\delta}^{nev}\right)  \right)  \cdot\dot{p}_{g}\left(  \pi
_{g}\right)  \right] \\
&  ~~~~~~~~-\mathbb{E}\left[  \alpha_{g}^{ps,nev}\left(  \pi_{g}\right)  \cdot
w_{g}^{comp}\left(  \pi_{g}\right)  \cdot\left(  Y_{t}-Y_{g-\delta
-1}-m_{g,t,\delta}^{nev}\left(  \beta_{g,t,\delta}^{nev}\right)  \right)
\cdot\dot{p}_{g}\left(  \pi_{g}\right)  \right]  ,
\end{align*}
with $\dot{m}_{g,t,\delta}^{nev}\left(  \beta_{g,t,\delta}^{nev}\right)
=\left.  \partial m_{g,t,\delta}^{nev}\left(  X;\beta_{g,t,\delta}%
^{nev}\right)  \right/  \partial\beta_{g,t,\delta}^{nev}$, $\dot{p}_{g}\left(
\pi_{g}\right)  =\left.  \partial p_{g}\left(  X;\pi_{g}\right)  \right/
\partial\pi_{g}$, and%
\[
\alpha_{g}^{ps,nev}\left(  \pi_{g}\right)  =\left.  \dfrac{C}{\left(
1-p_{g}\left(  X;\pi_{g}\right)  \right)  ^{2}}\right/  \mathbb{E}\left[
\dfrac{p_{g}\left(  X;\pi_{g}\right)  C}{1-p_{g}\left(  X;\pi_{g}\right)
}\right]  .
\]

Finally, let $ATT_{t\geq\left(  g-\delta\right)  }$ and $\widehat{ATT}%
_{t\geq\left(  g-\delta\right)  }^{dr,nev}$ denote the vector of $ATT(g,t)$
and $\widehat{ATT}_{dr}^{nev}(g,t;\delta)$, respectively, for all $g \in\mathcal{G}_\delta$, $t\in\left\{  2,\dots\mathcal{T-\delta
}\right\}  $ such that $t\geq g-\delta$. Analogously, let $\Psi_{t\geq\left(
g-\delta\right)  }^{dr,nev}$ denote the collection of $\psi_{g,t,\delta
}^{dr,nev}$ across all $g \in\mathcal{G}_\delta$,
$t\in\left\{  2,\dots\mathcal{T-\delta}\right\}  $ such that $t\geq g-\delta$.
Consider the following claim:%
\begin{align}
&  \text{For each } g \in\mathcal{G}_\delta \text{, }%
t\in\left\{  2,\dots\mathcal{T-\delta}\right\}  \text{ such that }t\geq
g-\delta\text{, }\nonumber\\
&  \exists\pi_{g}^{\ast}\in\Theta^{ps}:P\left(  p_{g}(X;\pi_{g}^{\ast}%
)=p_{g}\left(  X\right)  \right)  =1\text{ ~~or}\label{eqn:or_Ok}\\
&  \exists\beta_{g,t,\delta}^{\ast,nev}\in\Theta^{reg}:P\left(  m_{g,t,\delta
}^{nev}\left(  X;\beta_{g,t,\delta}^{\ast,nev}\right)  =m_{g,t,\delta}%
^{nev}\left(  X\right)  \right)  =1.\nonumber
\end{align}
Claim (\ref{eqn:or_Ok}) says that either the working parametric model for the
generalized propensity score is correctly specified, or the working outcome
regression model for the comparison group is correctly specified.

The next theorem establishes the joint limiting distribution of $\widehat{ATT}%
_{t\geq\left(  g-\delta\right)  }^{dr,nev}$.

\begin{theorem}
\label{thm:2} Under Assumptions \ref{ass:irrev}-\ref{ass:common-trends},
\ref{ass:overlap}-\ref{ass:integrabilitity}, for each $g$ and $t$ such that
$g\in\mathcal{G}_{\delta}$, $t\in\left\{  2,\dots\mathcal{T-\delta}\right\}  $
and $t\geq g-\delta$, provided that (\ref{eqn:or_Ok}) is true,%
\[
\sqrt{n}(\widehat{ATT}_{dr}^{nev}\left(  g,t;\delta\right)  -ATT\left(
g,t\right)  )=\frac{1}{\sqrt{n}}\sum_{i=1}^{n}\psi_{g,t,\delta}^{dr,nev}%
(W_{i};\kappa_{g,t}^{\ast,nev})+o_{p}(1).
\]
Furthermore, as $n\rightarrow\infty$,%
\[
\sqrt{n}(\widehat{ATT}_{t\geq\left(  g-\delta\right)  }^{dr,nev}%
-ATT_{t\geq\left(  g-\delta\right)  })\xrightarrow{d}N(0,\Sigma)
\]
where $\Sigma=\mathbb{E}[\Psi_{t\geq\left(  g-\delta\right)  }^{dr,nev}%
(W)\Psi_{t\geq\left(  g-\delta\right)  }^{dr,nev}(W)^{\prime}]$.
\end{theorem}

Theorem \ref{thm:2} provides the influence function for estimating the vector
of group-time average treatment effects, $ATT_{t\geq\left(  g-\delta\right)
},$ as well as its limiting distribution. Importantly, Theorem \ref{thm:2}
emphasizes the DR property of $\widehat{ATT}_{dr}^{nev}\left(  g,t;\delta
\right)  $: it recovers the $ATT\left(  g,t\right)  $ provided that either the
propensity score working model or outcome regression working model for the
\textquotedblleft never treated\textquotedblright\ is correctly specified.

In order to conduct inference, one can show that the sample analogue of
$\Sigma$ is a consistent estimator for $\Sigma$, which leads directly to
standard errors and pointwise confidence intervals. Instead of following this
route, we propose to use a simple multiplier bootstrap procedure to conduct
asymptotically valid inference. Our proposed bootstrap leverages the
asymptotic linear representations derived in Theorem \ref{thm:2} and inherits
important advantages. First, it is easy to implement and very fast to compute.
Each bootstrap iteration simply amounts to \textquotedblleft
perturbing\textquotedblright\ the influence function by a random weight $V$,
and it does not require re-estimating the propensity score in each bootstrap
draw. Second, in each bootstrap iteration, there are always observations from
each group. This can be a real problem with the traditional empirical
bootstrap where there may be no observations from a particular group in some
particular bootstrap iteration. Third, computation of simultaneously (in $g$
and $t$) valid confidence bands is relatively straightforward. This is
particularly important since researchers are likely to use confidence bands to
visualize estimation uncertainty about $ATT\left(  g,t\right)  .$ Unlike
pointwise confidence bands, simultaneous confidences bands do not suffer from
multiple-testing problems and are guaranteed to cover all $ATT\left(
g,t\right)$'s with a probability at least $1-\alpha$. Finally, we note that
our proposed bootstrap procedure can be readily modified to account for
clustering, see Remark \ref{rm:cluster} below.

To proceed, let $\widehat{\Psi}_{t\geq\left(  g-\delta\right)  }^{dr,nev}(W)$
denote the sample-analogue of $\Psi_{t\geq\left(  g-\delta\right)  }%
^{dr,nev}(W),$ where population expectations are replaced by their empirical
analogue, and the true nuisance functions and their derivatives are replaced
by their estimators. Let $\{V_{i}\}_{i=1}^{n}$ be a sequence of $iid$ random
variables with zero mean, unit variance, and finite third moment, independent
of the original sample $\{W_{i}\}_{i=1}^{n}$. A popular example involves $iid$
Bernoulli variates $\left\{  V_{i}\right\}  $ with $P\left(  V=1-\kappa
\right)  =\kappa/\sqrt{5}$ and $P\left(  V=\kappa\right)  =1-\kappa/\sqrt{5}$,
where $\kappa=\left(  \sqrt{5}+1\right)  /2,$ as suggested by
\cite{Mammen1993}.

We define $\widehat{ATT}_{t\geq\left(  g-\delta\right)  }^{\ast,dr,nev}$, a
bootstrap draw of $\widehat{ATT}_{t\geq\left(  g-\delta\right)  }^{dr,nev}$,
via
\begin{equation}
\widehat{ATT}_{t\geq\left(  g-\delta\right)  }^{\ast,dr,nev}=\widehat{ATT}%
_{t\geq\left(  g-\delta\right)  }^{dr,nev}+\mathbb{E}_{n}\left[
V\cdot\widehat{\Psi}_{t\geq\left(  g-\delta\right)  }^{dr,nev}(W)\right]  .
\label{eqn:att.boot}%
\end{equation}
The next theorem establishes the asymptotic validity of the multiplier
bootstrap procedure proposed above.

\begin{theorem}
\label{thm:bootstrap} Under the assumptions of Theorem \ref{thm:2}
\[
\sqrt{n}\left(  \widehat{ATT}_{t\geq\left(  g-\delta\right)  }^{\ast
,dr,nev}-\widehat{ATT}_{t\geq\left(  g-\delta\right)  }^{dr,nev}\right)
\underset{\ast}{\overset{d}{\rightarrow}}N(0,\Sigma),
\]
where $\Sigma$ is as in Theorem \ref{thm:2}, and $\underset{\ast
}{\overset{d}{\rightarrow}}$ denotes weak convergence (convergence in
distribution) of the bootstrap law in probability, i.e., conditional on the
original sample $\{W_{i}\}_{i=1}^{n}$. Additionally, for any continuous
functional $\Gamma(\cdot),$\footnote{Since the number of periods $\mathcal{T}$ is fixed, $\Gamma(\cdot)$ should be interpreted as a continuous functional between Euclidean spaces.}

\[
\Gamma\left(  \sqrt{n}\left(  \widehat{ATT}_{t\geq\left(  g-\delta\right)
}^{\ast,dr,nev}-\widehat{ATT}_{t\geq\left(  g-\delta\right)  }^{dr,nev}%
\right)  \right)  \underset{\ast}{\overset{d}{\rightarrow}}\Gamma\left(
N(0,\Sigma)\right)  .
\]

\end{theorem}

We now describe a practical bootstrap algorithm to compute studentized
confidence bands that cover $ATT\left(  g,t\right)  $ simultaneously over all
$t\geq g-\delta$ with a pre-specified probability $1-\alpha$ in large samples.
This is similar to the bootstrap procedures used in \cite{Kline2012},
\cite{Belloni2017} and \cite{chernozhukov2018sorted} in different contexts.

\begin{algorithm}
\label{alg:boot1} $1)$ Draw a realization of $\{V_{i}\}_{i=1}^{n}.~2)$ Compute
$\widehat{ATT}_{t\geq\left(  g-\delta\right)  }^{\ast,dr,nev}$ as in
(\ref{eqn:att.boot}), denote its $\left(  g,t\right)  $-element as
$\widehat{ATT}^{\ast}\left(  g,t\right)  ,$ and form a bootstrap draw of its
limiting distribution as
\[
\hat{R}^{\ast}\left(  g,t\right)  =\sqrt{n}\left(  \widehat{ATT}^{\ast}\left(
g,t\right)  -\widehat{ATT}\left(  g,t\right)  \right)  .
\]
$3)$ Repeat steps 1-2 $B$ times. $4)$ Compute a bootstrap estimator of the
main diagonal of $\Sigma^{1/2}$ such as the bootstrap interquartile range
normalized by the interquartile range of the standard normal distribution,
$\widehat{\Sigma}^{1/2}\left(  g,t\right)  =\left(  q_{0.75}\left(
g,t\right)  -q_{0.25}\left(  g,t\right)  \right)  /\left(  z_{0.75}%
-z_{0.25}\right)  ,$ where $q_{p}\left(  g,t\right)  $ is the $pth$ sample
quantile of the $\hat{R}^{\ast}\left(  g,t\right)  $ in the $B$ draws, and
$z_{p}$ is the $pth$ quantile of the standard normal distribution. $5)$ For
each bootstrap draw, compute $t\mathit{-}test_{t\geq\left(  g-\delta\right)
}=\max_{\left(  g,t\right)  }\left\vert \hat{R}^{\ast}\left(  g,t\right)
\right\vert \widehat{\Sigma}\left(  g,t\right)  ^{-1/2}.$ $5)$ Construct
$\widehat{c}_{1-\alpha}$ as the empirical $\left(  1-a\right)  $-quantile of
the $B$ bootstrap draws of $t\mathit{-}test_{t\geq\left(  g-\delta\right)  }$.
$6)$ Construct the bootstrapped simultaneous confidence band for $ATT\left(
g,t\right)  $, $t\geq\left(  g-\delta\right)  ,$ as $\widehat{C}\left(
g,t\right)  =[\widehat{ATT}_{dr}^{nev}\left(  g,t;\delta\right)
\pm\widehat{c}_{1-\alpha}\widehat{\Sigma}\left(  g,t\right)  ^{-1/2}/\sqrt
{n}].$
\end{algorithm}

The next corollary to Theorem \ref{thm:bootstrap} states that the simultaneous
confidence band for $ATT\left(  g,t\right)  $ described in Algorithm
\ref{alg:boot1} has correct asymptotic coverage.

\begin{corollary}
\label{cor:boot}Under the assumptions of Theorem \ref{thm:2}, for any
$0<\alpha<1,$ as $n\rightarrow\infty$,
\[
P\left(  ATT\left(  g,t\right)  \in\widehat{C}\left(  g,t\right)  ~~~\forall
t\in\left\{  2,\dots,\mathcal{T}\right\}  ,g\in\mathcal{G}_{\delta}:t\geq
g-\delta\right)  \rightarrow
1-\alpha,
\]
where $\widehat{C}\left(  g,t\right)  $ is as defined in Algorithm
\ref{alg:boot1}.
\end{corollary}

\begin{remark}
\label{rm:cluster}In DiD applications, it is common to use \textquotedblleft
cluster-robust\textquotedblright\ inference procedures; see, e.g.,
\cite{Wooldridge2003} and \cite{Bertrand2004}. However, we note that the
choice of whether to cluster or not is usually not obvious, and depends on the
kind of uncertainty one is trying to reflect; see, e.g., \cite{Abadie2017a}
for a discussion in a cross-sectional setup.\footnote{The formal results in
\cite{Abadie2017a} focus on the cross section case and rely on additional
functional form restrictions that we do not impose in this paper. Fully extending
the results of \cite{Abadie2017a} to the semiparametric panel data case is
beyond the scope of our paper.} In the case that one wishes to account for
clustering to reflect \textquotedblleft cluster-based\textquotedblright%
\ sampling uncertainty, we note that this can be done in a straightforward
manner using a small modification of the multiplier bootstrap described above,
provided that the number of cluster is \textquotedblleft
large.\textquotedblright\ More precisely, instead of drawing
observation-specific $V$'s, one simply needs to draw cluster-specific $V$'s;
see, e.g., \cite{Sherman2007a}, \cite{Kline2012}, \cite{Cheng2013}, and
\cite{Mackinnon2017,mackinnon2020randomization}. If the number of clusters is
\textquotedblleft small,\textquotedblright\ however, the application of the
aforementioned bootstrap procedure is not warranted.\footnote{In such cases,
provided that one is comfortable imposing additional functional form
assumptions, one could use alternative procedures such as \cite{Conley2011}
and \cite{Ferman2018}. Extending these proposals to our setup is beyond the
scope of this paper though.}
\end{remark}

\begin{remark}
\label{rm:studentization} In Algorithm \ref{alg:boot1} we have required an
estimator for the main diagonal of $\Sigma$. However, we note that if one
takes $\widehat{\Sigma}\left(  g,t\right)  =1$ for all $\left(  g,t\right)  $,
the result in Corollary \ref{cor:boot} continues to hold. However, the
resulting \textquotedblleft constant width\textquotedblright\ simultaneous
confidence band may be of larger length; see, e.g., \cite{MontielOlea2017} and
\cite{Freyberger2018}.
\end{remark}

\begin{remark}
 The above results focus on making inference about $ATT(g,t)$'s in (effective) post-treatment periods $t \ge g - \delta$. Although the limited anticipation condition in Assumption \ref{ass:anticipation} implies that $ATT(g,t)=0$ for all $t < g - \delta$  regardless of the group $g$, it is common practice to also estimate these pre-treatment parameters and use them to assess the credibility of the underlying identifying assumptions. Note that our DiD estimands (\ref{eqn:att_ipw_nev}) - (\ref{eqn:att_dr_ny}) can be easily adjusted to include these by simply replacing the ``long differences'' $(Y_{t} - Y_{g-\delta-1})$ with the ``short differences''  $(Y_{t} - Y_{t-1})$ for all $t < g - \delta$. All our results continues to hold when one augments $\widehat{ATT}_{t\geq\left(  g-\delta\right)  }^{dr,nev}$ to also include these estimates for the $ATT(g,t)$'s in the pre-treatment periods $t < g - \delta$.
\end{remark}

\subsection{Asymptotic Theory for Summary Parameters}

Assume, for simplicity, that Assumption \ref{ass:anticipation} holds with
$\delta=0$. In this section, we discuss how one can estimate and make
inference about the summary measures of the casual effects discussed in
Section \ref{sec:4}. More concisely, we consider parameters of the form of
$\theta$ as defined in (\ref{eq:aggte}), which covers all of the aggregated
parameters discussed in Section \ref{sec:4}.

Given the discussion in Section \ref{sec:asy-agg}, a natural way to estimate
$\theta$ is to use the plug-in type estimators
\[
\hat{\theta}=\sum_{g\in\mathcal{G}}\sum_{t=2}^{\mathcal{T}}\widehat{w}\left(
g,t\right)  \widehat{ATT}_{dr}^{nev}\left(  g,t;0\right)  ,
\]
where $\widehat{w}\left(  g,t\right)  $ are estimators for $w\left(
g,t\right)  $ such that for all $g\in\mathcal{G}$ and $t=2,\dots,\mathcal{T}%
$,
\[
\sqrt{n}\left(  \widehat{w}\left(  g,t\right)  -w\left(  g,t\right)  \right)
=\frac{1}{\sqrt{n}}\sum_{i=1}^{n}\xi_{g,t}^{w}(\mathcal{W}_{i})+o_{p}\left(
1\right)  ,
\]
with $\mathbb{E}\left[  \xi_{gt}^{w}(\mathcal{W})\right]  =0$ and
$\mathbb{E}\left[  \xi_{gt}^{w}(\mathcal{W})\xi_{gt}^{w}(\mathcal{W})^{\prime
}\right]  $ finite and positive definite. Estimators based on the sample
analogue of the weights discussed in Section \ref{sec:4} satisfy this condition.

Let%
\[
l^{w}\left(  W_{i}\right)  =\sum_{g\in\mathcal{G}}\sum_{t=2}^{\mathcal{T}%
}\left(  w\left(  g,t\right)  \cdot\psi_{g,t,0}^{dr,nev}(W_{i};\kappa
_{g,t}^{\ast,nev})+\xi_{g,t}^{w}(W_{i})\cdot ATT(g,t)\right)  ,
\]
where $\psi_{g,t,\delta}^{dr,nev}$ are as defined in (\ref{eqn:psigt}).

The following result follows immediately from Theorem \ref{thm:2}, and can be
used to conduct asymptotically valid inference for the summary causal
parameters $\theta$.

\begin{corollary}
\label{cor:1} Under the assumptions of Theorem \ref{thm:2},%
\begin{align*}
\sqrt{n}(\hat{\theta}-\theta)  &  =\frac{1}{\sqrt{n}}\sum_{i=1}^{n}%
l^{w}\left(  W_{i}\right)  +o_{p}(1)\\
&  \xrightarrow{d}N\left(  0,\mathbb{E}\left[  l^{w}\left(  W\right)
^{2}\right]  \right)
\end{align*}

\end{corollary}

Corollary \ref{cor:1} implies that one can construct standard errors and
confidence intervals for summary treatment effect parameters based on a
consistent estimator of $\mathbb{E}\left[  l^{w}\left(  W\right)  ^{2}\right]
$ or by using a bootstrap procedure like the one in Algorithm \ref{alg:boot1}.
The main advantage of using the bootstrap procedure akin to Algorithm
\ref{alg:boot1} is that inference procedures would be robust against
multiple-testing problems. This is particularly attractive when considering
$\theta_{es}(e)$, $\theta_{es}^{bal}(e;e^{\prime})$, $\theta_{sel}(\tilde{g}%
)$, and $\theta_{c}\left(  \tilde{t}\right)  $, as practitioners would
probably analyze how these parameters differ across event-times $e$, groups
$\tilde{g}$, and calendar-time $\tilde{t}$.

\begin{remark}
\label{rm:cluster2} As discussed in Remark \ref{rm:cluster}, the validity of
the \textquotedblleft cluster-robust\textquotedblright\ multiplier bootstrap
procedure relies on the number of clusters being \textquotedblleft
large.\textquotedblright\ In some applications such a condition may be more
plausible when analyzing the aggregated parameter $\theta$ than when analyzing
the $ATT(g,t)$ themselves.
\end{remark}

\section{The Effect of Minimum Wage Policy on Teen Employment}

In this section, we illustrate the empirical relevance of our proposed
methods. To do this, we apply our methods to study the effect of the minimum
wage on teen employment. The main goal of this section is to compare results
arising from using a TWFE specification (as is most common in applications) to
results coming from our proposed method. We think that this comparison is
important in order to get a sense of whether the theoretical limitations of
TWFE discussed in recent work end up translating into meaningful differences
in applications. Moreover, one might expect that understanding the effect of a
minimum wage change on employment is a challenging case for TWFE as the effect
of the minimum wage may be dynamic (\cite{Meer2016}) and the timing of minimum
wage changes varies across states. Unlike TWFE, the approach that we have
proposed in the current paper is robust to these challenges.

By far the most common approach to trying to understand the effect of the
minimum wage on employment is to exploit variation in the timing of minimum
wage increases across states. Our identification strategy follows this
approach. In particular, we consider a time period from 2001-2007 where the
federal minimum wage was flat at \$5.15 per hour. We focus on county level
teen employment in states whose minimum wage was equal to the federal minimum
wage at the beginning of the period. Some of these states increased their
minimum wage over this period -- these become treated groups.  In particular, we define groups by the time period when a state first increased its minimum wage.  Others did not increase their minimum wage -- these are the untreated group. This setup allows us to have more data than
local case study approaches. On the other hand, it also allows us to have
cleaner identification (state-level minimum wage policy changes) than in
studies with more periods; the latter setup is more complicated than ours
particularly because of the variation in the federal minimum wage over time.
It also allows us to check for internal consistency of identifying assumptions
-- namely whether or not the identifying assumptions hold in periods before
particular states raised their minimum wages.

We use county level data on teen employment and other county characteristics.
County level teen employment comes from the Quarterly Workforce Indicators
(QWI), as in \cite{Dube2016}; see \cite{Dube2016} for a detailed discussion of
this dataset. Other pre-treatment county characteristics come from the 2000
County Data Book. These include county population in 2000, the fraction of the
population that is white,
educational characteristics from 1990, median income in 1997, and the fraction
of the population below the poverty level in 1997.  After dropping ten states due to their minimum wage being higher than the federal minimum wage in 2000, seven other states for lack of data on teen employment, and four other states in the Northern census region, our final sample includes county-level data from 29 states.  We provide additional details on constructing the data in the Supplementary Appendix.

Summary statistics for county characteristics are provided in \Cref{tab:ss}.
There are some notable differences in county characteristics between counties
in states that increased their minimum wage and in states that did not
increase their minimum wage. Treated counties are much less likely to be in
the South. They also have much higher population (on average 94,000 compared
to 53,000 for untreated counties). The proportion of white residents is higher
in treated counties (on average, 89\% compared to 83\% for untreated
counties). There are smaller differences in the fraction with high school
degrees and the poverty rate though the differences are both statistically
significant. Treated counties have a somewhat higher fraction of high school
graduates and a somewhat lower poverty rate.

{ \singlespace
\newcolumntype{.}{D{.}{.}{-1}}
\ctable[caption={Summary Statistics for Main Dataset},label=tab:ss, pos=t,notespar]{lcccc}{\tnote[]{\textit{Notes}: Summary statistics for counties located in states that raised their minimum wage between Q2 of 2003 and Q1 of 2007 (treated) and states whose minimum wage was effectively set at the federal minimum wage for the entire period (untreated).  The sample consists of 2284 counties.
\textit{Sources: } Quarterly Workforce Indicators and 2000 County Data Book}}{\FL
\multicolumn{1}{l}{}&\multicolumn{1}{c}{Treated Counties}&\multicolumn{1}{c}{Untreated Counties}&\multicolumn{1}{c}{Diff.}&\multicolumn{1}{c}{P-val on Diff.}\ML
~~Midwest&0.59&0.34&0.25&0.00\NN
~~South&0.27&0.59&-0.32&0.00\NN
~~West&0.14&0.07&0.07&0.00\NN
~~Population (1000s)&94.32&53.43&40.89&0.00\NN
~~White&0.89&0.83&0.06&0.00\NN
~~HS Graduates&0.59&0.55&0.04&0.00\NN
~~Poverty Rate&0.13&0.16&-0.03&0.00\NN
~~Median Inc. (1000s)&33.91&31.89&2.02&0.00\LL
}}

\subsection{Results}

In the following we discuss different sets of results using different
identification strategies. In particular, we consider the cases in which one
would assume that the parallel trends assumption would hold unconditionally,
and when it holds only after controlling on observed characteristics $X$.  In the main text, we consider the case where never-treated counties are the comparison group and where we do not allow for any anticipation effects (i.e., $\delta=0$).  We provide results using the not-yet-treated counties as the comparison group and allowing for one year anticipation in the Supplementary Appendix; results from those cases are quite similar to the ones presented here.

\begin{figure}[t]
\caption{Minimum Wage Group-Time Average Treatment Effects}%
\label{fig:attgt}
\centering
\includegraphics[width=.9\textwidth, keepaspectratio]{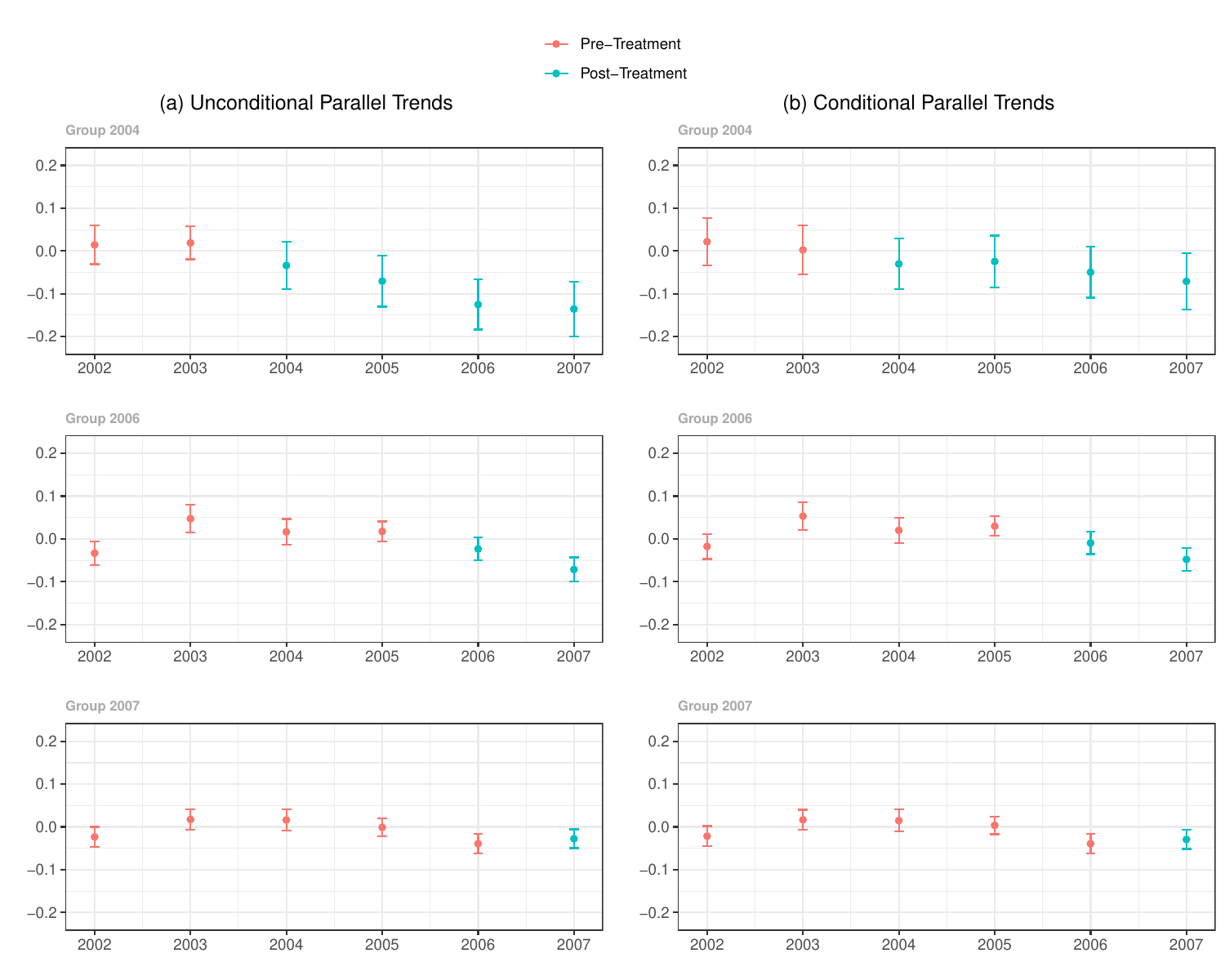}
\subcaption*{\scriptsize \textit{Notes:} The effect of the minimum wage on
teen employment estimated under the unconditional parallel trends assumption
(Panel (a)) and the conditional parallel trends assumption (Panel (b)). Red
lines give point estimates and uniform 95\% confidence bands for pre-treatment
periods allowing for clustering at the county level. Under the null hypothesis
of the parallel trends assumption holding in all periods, these should be
equal to 0. Blue lines provide point estimates and uniform 95\% confidence
bands for the treatment effect of increasing the minimum wage allowing for
clustering at the county level. The top row includes states that increased
their minimum wage in 2004, the middle row includes states that increased
their minimum wage in 2006, and the bottom row includes states that increased
their minimum wage in 2007. The estimates in Panel (b) use the the doubly robust estimator discussed in the text.}
\end{figure}

The first set of results comes from using the unconditional parallel trends
assumption to estimate the effect of raising the minimum wage on teen
employment. The results for group-time average treatment effects are reported
in Panel (a) of \Cref{fig:attgt} along with a uniform 95\% confidence band.
All inference procedures use clustered bootstrapped standard errors at the
county level, and account for the autocorrelation of the data. The plot
contains pre-treatment estimates that can be used to ``pre-test'' the parallel
trends assumption as well as treatment effect estimates in post-treatment periods.

The group-time average treatment effect estimates provide support for the view
that increasing the minimum wage led to a reduction in teen employment. For 5
out of 7 group-time average treatment effects, there is a clear statistically
significant negative effect on employment. The other two are marginally
insignificant (and negative). The group-time average treatment effects range
from 2.3\% lower teen employment to 13.6\% lower teen employment. The simple
average (weighted only by group size) is 5.2\% lower teen employment, and the
average effect of a minimum wage increase across all groups that increased
their minimum wage (corresponding to an estimate of $\theta_{sel}^{O}$ above)
is 3.9\% lower teen employment (see  Panel (a) of \Cref{tab:aggte}). A two-way
fixed effects model with a post treatment dummy variable also provides similar
results, indicating 3.7\% lower teen employment due to increasing the minimum
wage. In light of the literature on the minimum wage these results are not
surprising as they correspond to the types of regressions that tend to find
that increasing the minimum wage decreases employment; see the discussion in
\cite{Dube2010}.

As in \cite{Meer2016}, there also appears to be a dynamic effect of increasing
the minimum wage. For Illinois (the only state in the group that first raised
its minimum wage in 2004), teen employment is estimated to be 3.4\% lower on average in 2004
than it would have been if the minimum wage had not been increased. In 2005,
teen employment is estimated to be 7.1\% lower; in 2006, 12.5\% lower; and in
2007, 13.6\% lower. For states first treated in 2006, there is a small effect
in 2006: 2.3\% lower teen employment; however, it is larger in 2007: 7.1\%
lower teen employment.

Panel (a) of \Cref{tab:aggte} reports aggregated treatment effect measures.
First, we consider how the effect of increasing the minimum changes by the
amount of time that the policy has been in place. These parameters paint
largely the same picture as the group-time average treatment effects. The
effect of increasing the minimum wage on teen employment appears to be
negative and increasing in magnitude the longer states are exposed to the
higher minimum wage. In particular, in the first year that a state increases
its minimum wage, teen employment is estimated to decrease by 2.7\%, in the
second year it is estimated to decrease by 7.1\%, in the third year by 12.5\%,
and in the fourth year by 13.6\%. Notice that the last two dynamic treatment
effect estimates are exactly the same as the estimates coming from Illinois
alone because Illinois is the only state that is treated for at least two
years. These results are robust to keeping the composition of groups constant
by ``balancing'' the groups across different lengths of exposure to the 
treatment (see the row in \Cref{tab:aggte} labeled `Event Study w/ Balanced
Groups'). When we restrict the sample to only include groups that had a
minimum wage increase for at least one full year (i.e., we keep groups 2004
and 2006 but not 2007), we estimate that the effect of increasing the minimum
wage on impact is 2.7\% lower teen employment and 7.1\% lower teen employment
one year after the increase.\footnote{Notice that these estimates are exactly
the same as in the first two periods for the dynamic treatment effect
estimates that do not hold the composition of groups constant across different
lengths of exposure. The reason that they are the same for initial exposure is
coincidental as the results holding group composition constant do not include
the group first treated in 2007 (the estimated effect of the minimum wage in
2007 for the group of states first treated in 2007 is 2.76\% lower teen
employment which just happens to correspond to the estimated effect for the balanced groups). On the other hand, for the second period, they correspond by
construction because both estimates only include the groups first treated in
2004 and 2006.}

\setlength{\tabcolsep}{7pt} \renewcommand{\arraystretch}{.7}

\newcolumntype{.}{D{.}{.}{-1}}

\ctable[caption={Minimum Wage Aggregated Treatment Effect Estimates},label=tab:aggte,pos=h!,notespar,doinside=\footnotesize,mincapwidth=.95\textwidth]{lcccccc}{\tnote[]{\scriptsize\textit{Notes:}  The table reports aggregated treatment effect parameters under the unconditional and conditional parallel trends assumptions and with clustering at the county level.  The row `TWFE' reports the coefficient on a post-treatment dummy variable from a two-way fixed effects regression.  The row `Simple Weighted Average' reports the weighted average (by group size) of all available group-time average treatment effects as in \Cref{eqn:avg-overall}.  The row `Group-Specific Effects' summarizes average treatment effects by the  timing of the minimum wage increase; here, $g$ indexes the year that a county is first treated.  The row `Event Study' reports average treatment effects by the length of exposure to the minimum wage increase; here, $e$ indexes the length of exposure to the treatment.  The row `Calendar Time Effects' reports average treatment effects by year; here, $t$ indexes the year.  The row `Event Study w/ Balanced Groups' reports average treatment effects by length of exposure using a fixed set of groups at all lengths of exposure; here, $e$ indexes the length of exposure and the sample consists of counties that have at least two years of exposure to minimum wage increases.  The column `Single Parameters' represents a further aggregation of each type of parameter, as discussed in the text.  The estimates in Panel (b) use the the doubly robust estimator discussed in the text.}}{\FL
\multicolumn{7}{l}{}\NN[2pt]
\multicolumn{7}{l}{\bfseries (a) Unconditional Parallel Trends}\NN[3pt]
\multicolumn{1}{l}{\bfseries }&\multicolumn{4}{c}{Partially Aggregated}&\multicolumn{1}{c}{\bfseries }&\multicolumn{1}{c}{Single Parameters}\NN
\cline{2-5} \cline{7-7} \NN[1pt]
TWFE&&&&&&-0.037\NN
&&&&&&(0.006)\NN[7pt]
Simple Weighted Average&&&&&&-0.052\NN
&&&&&&(0.006)\NN[7pt]
Group-Specific Effects &\underline{g=2004}&\underline{g=2006}&\underline{g=2007}&&&\NN
&-0.091&-0.047&-0.028&&&-0.039\NN
&(0.019)&(0.008)&(0.007)&&&(0.007)\NN
Event Study&\underline{e=0}&\underline{e=1}&\underline{e=2}&\underline{e=3}&&\NN
&-0.027&-0.071&-0.125&-0.136&&-0.090\NN
&(0.006)&(0.009)&(0.021)&(0.023)&&(0.013)\NN
Calendar Time Effects&\underline{t=2004}&\underline{t=2005}&\underline{t=2006}&\underline{t=2007}&&\NN
&-0.034&-0.071&-0.055&-0.050&&-0.052\NN
&(0.019)&(0.02)&(0.009)&(0.006)&&(0.013)\NN
Event Study &\underline{e=0}&\underline{e=1}&&&&\NN
w/ Balanced Groups&-0.027&-0.071&&&&-0.049\NN
&(0.009)&(0.009)&&&&(0.008)\NN[10pt]
\multicolumn{7}{l}{\bfseries (b) Conditional Parallel Trends}\NN[2pt]
\multicolumn{1}{l}{\bfseries }&\multicolumn{4}{c}{Partially Aggregated}&\multicolumn{1}{c}{\bfseries }&\multicolumn{1}{c}{Single Parameters}\NN
\cline{2-5} \cline{7-7}\NN[1pt]
TWFE&&&&&&-0.008\NN
&&&&&&(0.006)\NN[7pt]
Simple Weighted Average&&&&&&-0.033\NN
&&&&&&(0.007)\NN[7pt]
Group-Specific Effects&\underline{g=2004}&\underline{g=2006}&\underline{g=2007}&&&\NN
&-0.044&-0.029&-0.029&&&-0.031\NN
&(0.020)&(0.008)&(0.008)&&&(0.007)\NN
Event Study &\underline{e=0}&\underline{e=1}&\underline{e=2}&\underline{e=3}&&\NN
&-0.024&-0.041&-0.050&-0.071&&-0.046\NN
&(0.006)&(0.009)&(0.022)&(0.026)&&(0.013)\NN
Calendar Time Effects&\underline{t=2004}&\underline{t=2005}&\underline{t=2006}&\underline{t=2007}&&\NN
&-0.030&-0.025&-0.030&-0.049&&-0.033\NN
&(0.022)&(0.021)&(0.009)&(0.007)&&(0.012)\NN
Event Study&\underline{e=0}&\underline{e=1}&&&&\NN
w/ Balanced Groups&-0.016&-0.041&&&&-0.028\NN
&(0.010)&(0.009)&&&&(0.008)\LL
}

Our summary parameters aggregated by group and by calendar time are also
consistent with the idea that increasing the minimum wage had a negative
effect on county level teen employment relative to what would have happened in
the absence of the minimum wage increase.

The second set of results comes from using the conditional parallel trends
assumption; that is, we assume only that counties with \textit{the same
characteristics} would follow the same trend in teen employment in the absence
of treatment. The county characteristics that we use are region of the
country, county population, county median income, the fraction of the
population that is white, the fraction of the population with a high school
education, and the county's poverty rate. We use the doubly robust estimation
procedure discussed above. Thus, estimation requires a first step estimation
of the generalized propensity score and outcome regression discussed above.
For each generalized propensity score, we estimate a logit model that includes
each county characteristic along with quadratic terms for population and
median income.\footnote{Using the propensity score specification tests
proposed by \cite{SantAnna2018}, we fail to reject the null hypothesis that
these models are correctly specified at the usual significance levels.} For
the outcome regressions, we use the same specification for the covariates.

Before presenting these results, we note that our doubly robust estimation
procedure is not computationally demanding. Our estimates of group-time
average treatment effects in this section (across all groups and time periods
and including our multiplier bootstrap with 1000 iterations) run in 3.0 
seconds on a laptop with a 2.80-GHz Intel i5 processor with 8GB of RAM and
without using any parallel processing.

For comparison's sake, we first estimate the coefficient on a post-treatment
dummy variable in a model with unit fixed effects and region-year fixed
effects. This is very similar to one of the sorts of models that
\cite{Dube2010} finds to eliminate the correlation between the minimum wage
and employment. Like \cite{Dube2010}, using this specification, we find that
the estimated coefficient is small and not statistically different from 0.
However, one must have in mind that the approach we proposed in this article
is different from the two-way fixed effects regression. In particular, we
explicitly identify group-time average treatment effects for different groups
and different times, allowing for arbitrary treatment effect heterogeneity as
long as the conditional parallel trends assumption is satisfied. Thus, our
causal parameters have a clear interpretation. As pointed out by
\cite{Wooldridge2005}, \cite{Chernozhukov2013}, \cite{DeChaisemartin2016},
\cite{Borusyak2017}, \cite{Goodman-bacon2017} and \cite{sloczynski-2018}, the
same may not be true for two-way fixed effects regressions in the presence of
treatment effect heterogeneity.\footnote{Our approach is also different from
that of \cite{Dube2010} in several other ways that are worth mentioning. We
focus on teen employment; \cite{Dube2010} considers employment in the
restaurant industry. Their most similar specification to the one mentioned
above includes census division-time fixed effects rather than region-time
fixed effects though the results are similar. Finally, our period of analysis
is different from theirs; in particular, there are no federal minimum wage
changes over the periods we analyze.}

The results using our approach are available in Panel (b) in \Cref{fig:attgt}
and Panel (b) in \Cref{tab:aggte}. Interestingly, we find quite different
results using our approach than are suggested by the two-way fixed effects
regression approach. In particular, we continue to find evidence that
increasing the minimum wage tended to reduce teen employment. The estimated
group-time average treatment effects range from 0.9\% lower teen employment
(not statistically different from 0) in 2006 for the group of states first
treated in 2006 to 7.1\% lower teen employment in 2007 for states first
treated in 2004. Now, 3 of 7 group-time average treatment effects are
statistically significant. The average effect of increasing the minimum wage
on teen employment across all groups that increased their minimum wage is a
3.1\% reduction in teen employment. This estimate is much different from the
TWFE estimate. In addition, the pattern of dynamic treatment effects where the
magnitude of the effect of increasing the minimum wage tends to increase with
length of exposure is the same as in the unconditional case.

Overall, our results suggest that increasing the minimum wage decreased teen
employment relative to what it would have been without the policy change.
However, there are some important limitations of our application. First, some
of the estimates of pseudo group-time average treatment effects in
pre-treatment periods in \Cref{fig:attgt} are significantly different from
zero which provides some suggestive evidence against the parallel trends
assumption.
Second, as discussed in the Supplementary Appendix, there is some heterogeneity in the size of the
minimum wage increase itself across states which could complicate the
interpretation of our results. Together, these suggest that our results should
be interpreted with some caution. That being said, we think that the key
takeaway from the application is that, (implicitly) holding the main
identifying assumptions constant, in a prominent application in economics that
has many very common features (treatment effect heterogeneity, dynamic
effects, and staggered treatment adoption) the choice of estimation method can
potentially lead to qualitatively different conclusions.

\section{Conclusion}

This paper has considered Difference-in-Differences methods in the case where
there are more than two time periods and units can become treated at different
points in time -- a commonly encountered setup in empirical work in economics. 
In this setup, we have proposed group-time average treatment effects, $ATT(g,t)$, that are the average treatment effect in period $t$ for the group of units first treated in period $g$. Unlike the more common approach of including a post-treatment dummy variable in a two-way fixed effects regression, $ATT(g,t)$ corresponds to a well defined treatment effect parameter. We also showed that once $ATT(g,t)$ has been obtained for different values of $g$ and $t$, they can be aggregated into other parameters to more concisely summarize heterogeneity with respect to some particular dimension of interest (such as length of exposure to the treatment) or, alternatively, into a single overall treatment effect parameter. In addition, our approach is suitable (i) for cases where the parallel trends assumption holds only after conditioning on covariates, (ii) using different comparison groups such as the never-treated or not-yet-treated, and (iii) when units can anticipate participating in the treatment and may adjust their behavior before the treatment is implemented.  We view such flexibility as an important component of our proposed methodology.

We also provided nonparametric identification results leading to outcome regression, inverse probability weighting, and doubly robust estimands.  Given that our nonparametric identification results are constructive, we proposed to estimate $ATT(g,t)$ using its sample analogue. We established consistency and asymptotic normality of the proposed estimators, and proved the validity of a powerful, but easy to implement, multiplier bootstrap procedure to construct simultaneous confidence bands for $ATT(g,t)$.  The computational costs of our approach are generally low, and code for implementing our approach is available in the R \texttt{did} package.

Finally, we applied our approach to study the effect of minimum wage increases on teen employment. We found some evidence that increasing the minimum wage led to reductions in teen employment.  More interestingly though, in some cases we found notable differences between the results coming from our approach relative to the more common two-way fixed effects approach.  These differences suggest that using an approach that is robust to treatment effect heterogeneity and dynamics should be strongly considered by applied researchers.

\section*{Appendix A: Proofs of Main Results\label{App.A}}

\renewcommand{\thesection}{A}
\renewcommand{\theassumption}{A.\arabic{assumption}}
\renewcommand{\thelemma}{A.\arabic{lemma}} \setcounter{lemma}{0}
\setcounter{equation}{0} \setcounter{figure}{0} \setcounter{section}{0}
\setcounter{assumption}{0} {\small We provide the proofs of our results in
this appendix. Before proceeding, we first state and prove several auxiliary
lemmas that help us to prove our main theorems. }

{\small Let
\[
ATT_{X}(g,t)=\mathbb{E}[Y_{t}(g)-Y_{t}(0)|X,G_{g}=1].
\]
}

\begin{lemma}
{\small \label{lem:att1} Let Assumptions \ref{ass:irrev}, \ref{ass:iid},
\ref{ass:anticipation}, \ref{ass:common-trends}, and \ref{ass:overlap} hold.
Then, for all $g$ and $t$ such that $g\in\mathcal{G}_{\delta}$, $t\in\left\{
2,\dots\mathcal{T-\delta}\right\}  $ and $t\geq g-\delta$,
\[
ATT_{X}(g,t)=\mathbb{E}[Y_{t}-Y_{g-\delta-1}|X,G_{g}=1]-\mathbb{E}%
[Y_{t}-Y_{g-\delta-1}|X,C=1]\ a.s..
\]
}
\end{lemma}

{\small \noindent\textbf{Proof of Lemma \ref{lem:att1}:} In what follows, take
all equalities to hold almost surely (a.s.). Then, we have that%
\begin{align*}
ATT_{X}(g,t)  &  =\mathbb{E}[Y_{t}\left(  g\right)  -Y_{g-\delta-1}\left(
0\right)  |X,G_{g}=1]-\mathbb{E}[Y_{t}\left(  0\right)  -Y_{g-\delta-1}\left(
0\right)  |X,G_{g}=1]\\
&  =\mathbb{E}[Y_{t}\left(  g\right)  -Y_{g-\delta-1}\left(  0\right)
|X,G_{g}=1]-\sum_{\ell=0}^{t-g-\delta}\mathbb{E}[\Delta Y_{t-\ell}\left(
0\right)  |X,G_{g}=1]\\
&  =\mathbb{E}[Y_{t}\left(  g\right)  -Y_{g-\delta-1}\left(  0\right)
|X,G_{g}=1]-\sum_{\ell=0}^{t-g-\delta}\mathbb{E}[\Delta Y_{t-\ell}\left(
0\right)  |X,C=1]\\
&  =\mathbb{E}[Y_{t}\left(  g\right)  -Y_{g-\delta-1}\left(  0\right)
|X,G_{g}=1]-\mathbb{E}[Y_{t}\left(  0\right)  -Y_{g-\delta-1}\left(  0\right)
|X,C=1]\\
&  =\mathbb{E}[Y_{t}-Y_{g-\delta-1}|X,G_{g}=1]-\mathbb{E}[Y_{t}-Y_{g-\delta
-1}|X,C=1
\end{align*}
where the first equality follows from adding and subtracting $\mathbb{E}%
[Y_{g-\delta-1}\left(  0\right)  |X,G_{g}=1]$, the second equality from simple
algebra, the third equality by Assumption \ref{ass:common-trends}, the fourth
equality by simple algebra, and the last equality from (\ref{eq:observational}%
) and Assumption \ref{ass:anticipation}. $\square\bigskip$ }

\begin{lemma}
{\small \label{lem:att1-ny} Let Assumptions \ref{ass:irrev}, \ref{ass:iid},
\ref{ass:anticipation}, \ref{ass:common-trends-ny} and \ref{ass:overlap} hold.
Then, for all $g$ and $t$ such that $g\in\mathcal{G}_{\delta}$, $t\in\left\{
2,\dots\mathcal{T-\delta}\right\}  $ with $g-\delta \le t < \bar{g}$
\[
ATT_{X}(g,t)=\mathbb{E}[Y_{t}-Y_{g-\delta-1}|X,G_{g}=1]-\mathbb{E}%
[Y_{t}-Y_{g-\delta-1}|X,D_{t+\delta}=0,G_{g}=0]\ a.s..
\]
}
\end{lemma}

{\small \noindent\textbf{Proof of Lemma \ref{lem:att1-ny}:} The proof follows
similar steps as the proof of Lemma \textbf{\ref{lem:att1}}. Taking all
equalities to hold almost surely (a.s.), we have that
\begin{align*}
ATT_{X}(g,t)  &  =\mathbb{E}[Y_{t}\left(  g\right)  -Y_{g-\delta-1}\left(
0\right)  |X,G_{g}=1]-\mathbb{E}[Y_{t}\left(  0\right)  -Y_{g-\delta-1}\left(
0\right)  |X,G_{g}=1]\\
&  =\mathbb{E}[Y_{t}\left(  g\right)  -Y_{g-\delta-1}\left(  0\right)
|X,G_{g}=1]-\sum_{\ell=0}^{t-g-\delta}\mathbb{E}[\Delta Y_{t-\ell}\left(
0\right)  |X,G_{g}=1]\\
&  =\mathbb{E}[Y_{t}\left(  g\right)  -Y_{g-\delta-1}\left(  0\right)
|X,G_{g}=1]-\sum_{\ell=0}^{t-g-\delta}\mathbb{E}[\Delta Y_{t-\ell}\left(
0\right)  |X,D_{t+\delta}=0,G_{g}=0]\\
&  =\mathbb{E}[Y_{t}\left(  g\right)  -Y_{g-\delta-1}\left(  0\right)
|X,G_{g}=1]-\mathbb{E}[Y_{t}\left(  0\right)  -Y_{g-\delta-1}\left(  0\right)
|X,D_{t+\delta}=0,G_{g}=0]\\
&  =\mathbb{E}[Y_{t}-Y_{g-\delta-1}|X,G_{g}=1]-\mathbb{E}[Y_{t}-Y_{g-\delta
-1}|X,D_{t+\delta}=0,G_{g}=0]
\end{align*}
where the first equality follows from adding and subtracting $\mathbb{E}%
[Y_{g-\delta-1}\left(  0\right)  |X,G_{g}=1]$, the second equality from simple
algebra, the third equality by Assumption \ref{ass:common-trends-ny} with
$s=t+\delta$, the fourth equality by simple algebra, and the last equality from
(\ref{eq:observational}) and Assumption \ref{ass:anticipation}. $\square
\bigskip$ }

{\small Now, we are ready to proceed with the proofs of our main
theorems.\bigskip}

{\small \noindent\textbf{Proof of Theorem \ref{thm:att}:} }

{\small \textbf{Part 1: Identification when Assumption \ref{ass:common-trends}
is invoked. } }

{\small In this case, given the result in Lemma \ref{lem:att1},
\begin{align*}
ATT(g,t)  &  =\mathbb{E}[ATT_{X}(g,t)|G_{g}=1]\\
&  =\mathbb{E}\left[  \left.  \left(  \mathbb{E}[Y_{t}-Y_{g-\delta-1}%
|X,G_{g}=1]-\mathbb{E}[Y_{t}-Y_{g-\delta-1}|X,C=1]\right)  \right\vert
G_{g}=1\right] \\
&  =\mathbb{E}[Y_{t}-Y_{g-\delta-1}|G_{g}=1]-\mathbb{E}\left[
\underset{=m_{g,t,\delta}^{nev}\left(  X\right)  }{\underbrace{\mathbb{E}%
[Y_{t}-Y_{g-\delta-1}|X,C=1]}}|G_{g}=1\right] \\
&  =\mathbb{E}\left[  \frac{G_{g}}{\mathbb{E}\left[  G_{g}\right]  }\left(
Y_{t}-Y_{g-\delta-1}-m_{g,t,\delta}^{nev}\left(  X\right)  \right)  \right]  .
\end{align*}
Hence, we have that $ATT(g,t)=ATT_{or}^{nev}\left(  g,t;\delta\right)  .$ }

{\small Next, to show that $ATT(g,t)=ATT_{ipw}^{nev}\left(  g,t;\delta\right)
$, it suffices to show that
\begin{equation}
\frac{\mathbb{E}\left[  \dfrac{p_{g}\left(  X\right)  C}{(1-p_{g}(X))}%
(Y_{t}-Y_{g-\delta-1})\right]  }{\mathbb{E}\left[  \dfrac{p_{g}\left(
X\right)  C}{(1-p_{g}(X))}\right]  }=\frac{\mathbb{E}\left[  G_{g}%
\cdot\mathbb{E}[Y_{t}-Y_{g-\delta-1}|X,C=1]\right]  }{\mathbb{E}\left[
G_{g}\right]  }. \label{eqn:ipw}%
\end{equation}
Towards this end, by noticing that%
\begin{equation}
p_{g}\left(  X\right)  =\frac{\mathbb{E}\left[  G_{g}|X\right]  }%
{\mathbb{E}\left[  G_{g}+C|X\right]  },~1-p_{g}\left(  X\right)
=\frac{\mathbb{E}\left[  C|X\right]  }{\mathbb{E}\left[  G_{g}+C|X\right]  },
\label{eqn:pscore_nev}%
\end{equation}
it follows that%
\begin{align}
\mathbb{E}\left[  \dfrac{p_{g}\left(  X\right)  C}{1-p_{g}(X)}\right]   &
=\mathbb{E}\left[  \dfrac{\mathbb{E}\left[  G_{g}|X\right]  C}%
{\mathbb{E}\left[  C|X\right]  }\right] \nonumber\\
&  =\mathbb{E}\left[  \dfrac{\mathbb{E}\left[  G_{g}|X\right]  \mathbb{E}%
\left[  C|X\right]  }{\mathbb{E}\left[  C|X\right]  }\right] \nonumber\\
&  =\mathbb{E}\left[  \mathbb{E}\left[  G_{g}|X\right]  \right] \nonumber\\
&  =\mathbb{E}\left[  G_{g}\right]  . \label{eqn:Hajek}%
\end{align}
Next, by exploiting (\ref{eqn:pscore_nev}) and applying the law of iterated
expectations, we have that%
\begin{align*}
\mathbb{E}\left[  \dfrac{p_{g}\left(  X\right)  C}{(1-p_{g}(X))}%
(Y_{t}-Y_{g-\delta-1})\right]   &  =\mathbb{E}\left[  \dfrac{\mathbb{E}\left[
G_{g}|X\right]  C}{\mathbb{E}\left[  C|X\right]  }(Y_{t}-Y_{g-\delta
-1})\right] \\
&  =\mathbb{E}\left[  \dfrac{\mathbb{E}\left[  G_{g}|X\right]  }%
{\mathbb{E}\left[  C|X\right]  }\mathbb{E}\left[  \left.  C\cdot
(Y_{t}-Y_{g-\delta-1})\right\vert X\right]  \right] \\
&  =\mathbb{E}\left[  \mathbb{E}\left[  G_{g}|X\right]  \cdot\mathbb{E}\left[
\left.  (Y_{t}-Y_{g-\delta-1})\right\vert X,C=1\right]  \right] \\
&  =\mathbb{E}\left[  G_{g}\cdot\mathbb{E}\left[  \left.  (Y_{t}%
-Y_{g-\delta-1})\right\vert X,C=1\right]  \right]  .
\end{align*}
Thus, combined this with (\ref{eqn:Hajek}), we establish (\ref{eqn:ipw}),
implying that $ATT(g,t)=ATT_{ipw}^{nev}\left(  g,t;\delta\right)  $. }

{\small Finally, notice that
\begin{align*}
ATT_{dr}^{nev}\left(  g,t;\delta\right)   &  =\mathbb{E}\left[  \left(
\frac{G_{g}}{\mathbb{E}\left[  G_{g}\right]  }-\frac{\dfrac{p_{g}\left(
X\right)  C}{1-p_{g}\left(  X\right)  }}{\mathbb{E}\left[  \dfrac{p_{g}\left(
X\right)  C}{1-p_{g}\left(  X\right)  }\right]  }\right)  \left(
Y_{t}-Y_{g-\delta-1}-m_{g,t,\delta}^{nev}\left(  X\right)  \right)  \right] \\
&  =\underset{\equiv ATT_{ipw}^{nev}\left(  g,t;\delta\right)
}{\underbrace{\mathbb{E}\left[  \left(  \frac{G_{g}}{\mathbb{E}\left[
G_{g}\right]  }-\frac{\dfrac{p_{g}\left(  X\right)  C}{1-p_{g}\left(
X\right)  }}{\mathbb{E}\left[  \dfrac{p_{g}\left(  X\right)  C}{1-p_{g}\left(
X\right)  }\right]  }\right)  \left(  Y_{t}-Y_{g-\delta-1}\right)  \right]  }%
}\\
&  ~~~~~~~~~~~~~~~-\mathbb{E}\left[  \left(  \frac{G_{g}}{\mathbb{E}\left[
G_{g}\right]  }-\frac{\dfrac{p_{g}\left(  X\right)  C}{1-p_{g}\left(
X\right)  }}{\mathbb{E}\left[  \dfrac{p_{g}\left(  X\right)  C}{1-p_{g}\left(
X\right)  }\right]  }\right)  m_{g,t,\delta}^{nev}\left(  X\right)  \right] \\
&  =ATT\left(  g,t\right)  -\frac{1}{\mathbb{E}\left[  G_{g}\right]
}\mathbb{E}\left[  \left(  G_{g}-\dfrac{\mathbb{E}\left[  G_{g}|X\right]
C}{\mathbb{E}\left[  C|X\right]  }\right)  m_{g,t,\delta}^{nev}\left(
X\right)  \right] \\
&  =ATT\left(  g,t\right)  -\frac{1}{\mathbb{E}\left[  G_{g}\right]
}\mathbb{E}\left[  \left(  \mathbb{E}\left[  G_{g}|X\right]  -\mathbb{E}%
\left[  G_{g}|X\right]  \right)  \cdot m_{g,t,\delta}^{nev}\left(  X\right)
\right] \\
&  =ATT\left(  g,t\right)  .
\end{align*}
where the third equality follows from $ATT_{ipw}^{nev}\left(  g,t;\delta
\right)  =ATT\left(  g,t\right)  $, (\ref{eqn:pscore_nev}) and
(\ref{eqn:Hajek}), and the fourth equality from the law of iterated
expectations.\medskip}

{\small \textbf{Part 2: Identification when Assumption
\ref{ass:common-trends-ny} is invoked. } }

{\small In this case, given the result in Lemma \ref{lem:att1-ny},
\begin{align*}
ATT(g,t)  &  =\mathbb{E}[ATT_{X}(g,t)|G_{g}=1]\\
&  =\mathbb{E}\left[  \left.  \left(  \mathbb{E}[Y_{t}-Y_{g-\delta-1}%
|X,G_{g}=1]-\mathbb{E}[Y_{t}-Y_{g-\delta-1}|X,D_{t+\delta}=0,G_{g}=0]\right)
\right\vert G_{g}=1\right] \\
&  =\mathbb{E}[Y_{t}-Y_{g-\delta-1}|G_{g}=1]-\mathbb{E}\left[
\underset{=m_{g,t,\delta}^{ny}\left(  X\right)  }{\underbrace{\mathbb{E}%
[Y_{t}-Y_{g-\delta-1}|X,D_{t+\delta}=0,G_{g}=0]}}|G_{g}=1\right] \\
&  =\mathbb{E}\left[  \frac{G_{g}}{\mathbb{E}\left[  G_{g}\right]  }\left(
Y_{t}-Y_{g-1-a}-m_{g,t,\delta}^{ny}\left(  X\right)  \right)  \right]  .
\end{align*}
Hence, we have that $ATT(g,t)=ATT_{or}^{ny}\left(  g,t;\delta\right)  .$ }

{\small Next, to show that $ATT(g,t)=ATT_{ipw}^{ny}\left(  g,t;\delta\right)
$, it suffices to show that
\begin{equation}
\frac{\mathbb{E}\left[  \dfrac{p_{g,t+\delta}\left(  X\right)  \left(
1-D_{t+\delta}\right)  \left(  1-G_{g}\right)  }{1-p_{g,t+\delta}\left(
X\right)  }(Y_{t}-Y_{g-\delta-1})\right]  }{\mathbb{E}\left[  \dfrac
{p_{g,t+\delta}\left(  X\right)  \left(  1-D_{t+\delta}\right)  \left(
1-G_{g}\right)  }{1-p_{g,t+\delta}\left(  X\right)  }\right]  }=\frac
{\mathbb{E}\left[  G_{g}\cdot\mathbb{E}[Y_{t}-Y_{g-\delta-1}|X,D_{t+\delta
}=0,G_{g}=0]\right]  }{\mathbb{E}\left[  G_{g}\right]  }. \label{eqn:ipw_ny}%
\end{equation}
Towards this end, recall that $p_{g,t+\delta}(X)=P(G_{g}=1|X,G_{g}+\left(
1-D_{t+\delta}\right)  \left(  1-G_{g}\right)  =1)$ and also notice that 
\begin{equation}
p_{g,t+\delta}\left(  X\right)  =\frac{\mathbb{E}\left[  G_{g}|X\right]
}{\mathbb{E}\left[  G_{g}+\left(  1-D_{t+\delta}\right)  \left(
1-G_{g}\right)  |X\right]  },~1-p_{g}\left(  X\right)  =\frac{\mathbb{E}%
\left[  \left(  1-D_{t+\delta}\right)  \left(  1-G_{g}\right)  |X\right]
}{\mathbb{E}\left[  G_{g}+\left(  1-D_{t+\delta}\right)  \left(
1-G_{g}\right)  |X\right]  }, \label{eqn:pscore_ny}%
\end{equation}
it follows that by the law of iterated expectations,%
\begin{align}
\mathbb{E}\left[  \dfrac{p_{g,t+\delta}\left(  X\right)  \left(
1-D_{t+\delta}\right)  \left(  1-G_{g}\right)  }{1-p_{g,t+\delta}\left(
X\right)  }\right]   &  =\mathbb{E}\left[  \dfrac{\mathbb{E}\left[
G_{g}|X\right]  \left(  1-D_{t+\delta}\right)  \left(  1-G_{g}\right)
}{\mathbb{E}\left[  \left(  1-D_{t+\delta}\right)  \left(  1-G_{g}\right)
|X\right]  }\right] \nonumber\\
&  =\mathbb{E}\left[  G_{g}\right]  . \label{eqn:hajek_ny}%
\end{align}
Next, by exploiting (\ref{eqn:pscore_ny}) and applying the law of iterated
expectations, we have that%
\begin{align*}
&  \mathbb{E}\left[  \dfrac{p_{g,t+\delta}\left(  X\right)  \left(
1-D_{t+\delta}\right)  \left(  1-G_{g}\right)  }{1-p_{g,t+\delta}\left(
X\right)  }(Y_{t}-Y_{g-\delta-1})\right] \\
&  =\mathbb{E}\left[  \dfrac{\mathbb{E}\left[  G_{g}|X\right]  \left(
1-D_{t+\delta}\right)  \left(  1-G_{g}\right)  }{\mathbb{E}\left[  \left(
1-D_{t+\delta}\right)  \left(  1-G_{g}\right)  |X\right]  }(Y_{t}%
-Y_{g-\delta-1})\right] \\
&  =\mathbb{E}\left[  \dfrac{\mathbb{E}\left[  G_{g}|X\right]  }%
{\mathbb{E}\left[  \left(  1-D_{t+\delta}\right)  \left(  1-G_{g}\right)
|X\right]  }\mathbb{E}\left[  \left.  \left(  1-D_{t+\delta}\right)  \left(
1-G_{g}\right)  \cdot(Y_{t}-Y_{g-\delta-1})\right\vert X\right]  \right] \\
&  =\mathbb{E}\left[  \mathbb{E}\left[  G_{g}|X\right]  \cdot\mathbb{E}\left[
\left.  (Y_{t}-Y_{g-\delta-1})\right\vert X,D_{t+\delta}=0,G_{g}=0\right]
\right] \\
&  =\mathbb{E}\left[  G_{g}\cdot\mathbb{E}\left[  \left.  (Y_{t}%
-Y_{g-\delta-1})\right\vert X,D_{t+\delta}=0,G_{g}=0\right]  \right]  .
\end{align*}
Thus, combined this with (\ref{eqn:hajek_ny}), we establish (\ref{eqn:ipw_ny}%
), implying that $ATT(g,t)=ATT_{ipw}^{ny}\left(  g,t;\delta\right)  $. }

{\small Finally, notice that
\begin{align*}
ATT_{dr}^{ny}\left(  g,t;\delta\right)   &  =\mathbb{E}\left[  \left(
\frac{G_{g}}{\mathbb{E}\left[  G_{g}\right]  }-\frac{\dfrac{p_{g,t+\delta
}\left(  X\right)  \left(  1-D_{t+\delta}\right)  \left(  1-G_{g}\right)
}{1-p_{g,t+\delta}\left(  X\right)  }}{\mathbb{E}\left[  \dfrac{p_{g,t+\delta
}\left(  X\right)  \left(  1-D_{t+\delta}\right)  \left(  1-G_{g}\right)
}{1-p_{g,t+\delta}\left(  X\right)  }\right]  }\right)  \left(  Y_{t}%
-Y_{g-\delta-1}-m_{g,t,\delta}^{ny}\left(  X\right)  \right)  \right] \\
&  =ATT_{ipw}^{ny}\left(  g,t;\delta\right)  -\frac{1}{\mathbb{E}\left[
G_{g}\right]  }\mathbb{E}\left[  \left(  G_{g}-\dfrac{\mathbb{E}\left[
G_{g}|X\right]  \left(  1-D_{t+\delta}\right)  \left(  1-G_{g}\right)
}{\mathbb{E}\left[  \left(  1-D_{t+\delta}\right)  \left(  1-G_{g}\right)
|X\right]  }\right)  m_{g,t,\delta}^{ny}\left(  X\right)  \right] \\
&  =ATT\left(  g,t\right)  -\frac{1}{\mathbb{E}\left[  G_{g}\right]
}\mathbb{E}\left[  \left(  \mathbb{E}\left[  G_{g}|X\right]  -\mathbb{E}%
\left[  G_{g}|X\right]  \right)  \cdot m_{g,t,\delta}^{nev}\left(  X\right)
\right] \\
&  =ATT\left(  g,t\right)  .
\end{align*}
where the second equality follows from (\ref{eqn:pscore_ny}) and
(\ref{eqn:hajek_ny}), and the third equality from $ATT_{ipw}^{nev}\left(
g,t;\delta\right)  =ATT\left(  g,t\right)  $ and the law of iterated
expectations.$\square\bigskip$ }

{\small \textbf{Proof of Theorem \ref{thm:2}: }From Theorem
\textbf{\ref{thm:att}} it follows that $ATT\left(  g,t\right)  $'s are
point-identified for all groups $g$ and time periods $t$ such that
$g\in\mathcal{G}_{\delta}$, $t\in\left\{  2,\dots\mathcal{T-\delta}\right\}  $
and $t\geq g-\delta$. For each $(g,t)$ pair, the asymptotic linear
representation of $\sqrt{n}(\widehat{ATT}_{dr}^{nev}\left(  g,t;\delta\right)
-ATT\left(  g,t\right)  )$ follows from Theorem A.1(a) of \cite{SantAnna2020},
whereas
\[
\sqrt{n}(\widehat{ATT}_{t\geq\left(  g-\delta\right)  }^{dr,nev}%
-ATT_{t\geq\left(  g-\delta\right)  })\xrightarrow{d}N(0,\mathbb{E}%
[\Psi_{t\geq\left(  g-\delta\right)  }^{dr,nev}(W)\Psi_{t\geq\left(
g-\delta\right)  }^{dr,nev}(W)^{\prime}])
\]
follows from the Lindeberg--L\'{e}vy central limit theorem.$\square\bigskip$ }

{\small \noindent\textbf{Proof of Theorem \ref{thm:bootstrap}:} Note that, by
the conditional multiplier central limit theorem, see Lemma 2.9.5 in
\cite{VanderVaart1996}, as $n\rightarrow\infty$,%
\begin{equation}
\frac{1}{\sqrt{n}}\sum_{i=1}^{n}V_{i}\cdot\Psi_{t\geq\left(  g-\delta\right)
}^{dr,nev}(W_{i})\underset{\ast}{\overset{d}{\rightarrow}}N(0,\Sigma
),\label{eq:boot.0}%
\end{equation}
where $\Sigma=\mathbb{E}[\Psi_{t\geq\left(  g-\delta\right)  }^{dr,nev}%
(W)\Psi_{t\geq\left(  g-\delta\right)  }^{dr,nev}(W)^{\prime}].$ Thus, to
conclude the proof that
\[
\sqrt{n}\left(  \widehat{ATT}_{t\geq\left(  g-\delta\right)  }^{\ast
,dr,nev}-\widehat{ATT}_{t\geq\left(  g-\delta\right)  }^{dr,nev}\right)
\underset{\ast}{\overset{d}{\rightarrow}}N(0,\Sigma),
\]
it suffices to show that, for all $g$ and $t$ such that $g\in\mathcal{G}%
_{\delta}$, $t\in\left\{  2,\dots\mathcal{T-\delta}\right\}  $ and $t\geq
g-\delta$,
\begin{equation}
\frac{1}{\sqrt{n}}\sum_{i=1}^{n}V_{i}\cdot\left[  \widehat{\psi}_{g,t,\delta
}^{dr,nev}(W_{i};\widehat{\kappa}_{g,t}^{nev})-\psi_{g,t,\delta}%
^{dr,nev}(W_{i};\kappa_{g,t}^{\ast,nev})\right]  =o_{p^{\ast}}\left(
1\right)  ,\label{boot:needed}%
\end{equation}
where $\kappa_{g,t}^{\ast,nev}=\left(  \pi_{g}^{\ast\prime},\beta_{g,t,\delta
}^{\ast,nev~\prime}\right)  ^{\prime}$ is the vector of pseudo-true
finite-dimensional parameters. }

{\small Towards this end, recall that%
\[
\widehat{\psi}_{g,t,\delta}^{dr,nev}(W_{i};\widehat{\kappa}_{g,t}%
^{nev})=\widehat{\psi}_{g,t,\delta}^{treat,nev}(W_{i};\widehat{\beta
}_{g,t,\delta}^{nev})-\widehat{\psi}_{g,t,\delta}^{comp,nev}(W_{i}%
;\widehat{\pi}_{g},\widehat{\beta}_{g,t,\delta}^{nev})-\widehat{\psi
}_{g,t,\delta}^{est,nev}(W_{i};\widehat{\pi}_{g},\widehat{\beta}_{g,t,\delta
}^{nev})
\]
where
\begin{align*}
\widehat{\psi}_{g,t,\delta}^{treat,nev}(W;\widehat{\beta}_{g,t,\delta}^{nev})
&  =\widehat{w}_{g}^{treat}\cdot\left(  Y_{t}-Y_{g-\delta-1}-m_{g,t,\delta
}^{nev}\left(  \widehat{\beta}_{g,t,\delta}^{nev}\right)  \right) \\
&  ~~~~~~~~-\widehat{w}_{g}^{treat}\cdot\mathbb{E}_{n}\left[  \widehat{w}%
_{g}^{treat}\cdot\left(  Y_{t}-Y_{g-\delta-1}-m_{g,t,\delta}^{nev}\left(
\widehat{\beta}_{g,t,\delta}^{nev}\right)  \right)  \right]  ,\\
\widehat{\psi}_{g,t,\delta}^{comp,nev}(W;\widehat{\pi}_{g},\widehat{\beta
}_{g,t,\delta}^{nev})  &  =\widehat{w}_{g}^{comp,nev}\left(  \widehat{\pi}%
_{g}\right)  \cdot\left(  Y_{t}-Y_{g-\delta-1}-m_{g,t,\delta}^{nev}\left(
\widehat{\beta}_{g,t,\delta}^{nev}\right)  \right) \\
&  ~~~~~~~~-\widehat{w}_{g}^{comp}\left(  \widehat{\pi}_{g}\right)
\cdot\mathbb{E}_{n}\left[  w_{g}^{comp}\left(  \widehat{\pi}_{g}\right)
\cdot\left(  Y_{t}-Y_{g-\delta-1}-m_{g,t,\delta}^{nev}\left(  \widehat{\beta
}_{g,t,\delta}^{nev}\right)  \right)  \right]  ,\\
\widehat{\psi}_{g,t}^{est,nev}(W;\widehat{\pi}_{g},\widehat{\beta}%
_{g,t,\delta}^{nev})  &  =l_{g,t}^{or,nev}\left(  \widehat{\beta}_{g,t,\delta
}^{nev}\right)  ^{\prime}\cdot\widehat{M}_{g,t,\delta}^{dr,nev,1}%
+l_{g}^{ps,nev}\left(  \widehat{\pi}_{g}\right)  ^{\prime}\cdot\widehat{M}%
_{g,t,\delta}^{dr,nev,2},
\end{align*}
with%
\[
\widehat{w}_{g}^{treat}=\frac{G_{g}}{\mathbb{E}_{n}\left[  G_{g}\right]
},~~~~\widehat{w}_{g}^{comp,nev}\left(  \widehat{\pi}_{g}\right)
=\frac{\dfrac{\widehat{p}_{g}\left(  X;\widehat{\pi}_{g}\right)
C}{1-\widehat{p}_{g}\left(  X;\widehat{\pi}_{g}\right)  }}{\mathbb{E}%
_{n}\left[  \dfrac{\widehat{p}_{g}\left(  X;\widehat{\pi}_{g}\right)
C}{1-\widehat{p}_{g}\left(  X;\widehat{\pi}_{g}\right)  }\right]  },
\]
and%
\begin{align*}
\widehat{M}_{g,t,\delta}^{dr,nev,1}  &  =\mathbb{E}_{n}\left[  \left(
\widehat{w}_{g}^{treat}-\widehat{w}_{g}^{comp,nev}\left(  \widehat{\pi}_{g}\right)
\right)  \cdot\dot{m}_{g,t,\delta}^{nev}\left(  \widehat{\beta}_{g,t,\delta
}^{nev}\right)  \right]  ,\\
\widehat{M}_{g,t,\delta}^{dr,nev,2}  &  =\mathbb{E}_{n}\left[  \widehat{\alpha
}_{g}^{ps,nev}\left(  \widehat{\pi}_{g}\right)  \cdot\left(  Y_{t}%
-Y_{g-\delta-1}-m_{g,t,\delta}^{nev}\left(  \widehat{\beta}_{g,t,\delta}%
^{nev}\right)  \right)  \cdot\dot{p}_{g}\left(  \widehat{\pi}_{g}\right)
\right] \\
&  ~~~~~~~~-\mathbb{E}_{n}\left[  \widehat{\alpha}_{g}^{ps,nev}\left(
\widehat{\pi}_{g}\right)  \cdot\widehat{w}_{g}^{comp,nev}\left(  \widehat{\pi
}_{g}\right)  \cdot\left(  Y_{t}-Y_{g-\delta-1}-m_{g,t,\delta}^{nev}\left(
\widehat{\beta}_{g,t,\delta}^{nev}\right)  \right)  \cdot\dot{p}_{g}\left(
\widehat{\pi}_{g}\right)  \right]  ,\\
\widehat{\alpha}_{g}^{ps,nev}\left(  \widehat{\pi}_{g}\right)   &  =\left.
\dfrac{C}{\left(  1-p_{g}\left(  X;\widehat{\pi}_{g}\right)  \right)  ^{2}%
}\right/  \mathbb{E}_{n}\left[  \dfrac{p_{g}\left(  X;\widehat{\pi}%
_{g}\right)  C}{1-p_{g}\left(  X;\widehat{\pi}_{g}\right)  }\right]  .
\end{align*}
}

{\small We first show that
\begin{equation}
\frac{1}{\sqrt{n}}\sum_{i=1}^{n}V_{i}\cdot\left(  \widehat{\psi}_{g,t,\delta
}^{treat,nev}(W_{i};\widehat{\beta}_{g,t,\delta}^{nev})-\psi_{g,t,\delta
}^{treat,nev}(W_{i};\beta_{g,t,\delta}^{\ast,nev})\right)  =o_{p^{\ast}%
}\left(  1\right)  . \label{boot:treated}%
\end{equation}
Using the mean-value theorem, we write%
\begin{align*}
&  \frac{1}{\sqrt{n}}\sum_{i=1}^{n}V_{i}\cdot\widehat{\psi}_{g,t,\delta
}^{treat,nev}(W_{i};\widehat{\beta}_{g,t,\delta}^{nev})\\
&  =\frac{1}{\sqrt{n}}\sum_{i=1}^{n}V_{i}\cdot\widehat{w}_{g,i}^{treat}%
\cdot\left(  Y_{i,t}-Y_{i,g-\delta-1}-m_{g,t,\delta}^{nev}\left(  W_{i}%
;\beta_{g,t,\delta}^{\ast,nev}\right)  \right) \\
&  -\sqrt{n}\left(  \widehat{\beta}_{g,t,\delta}^{nev}-\beta_{g,t,\delta
}^{\ast,nev}\right)  ^{\prime}\frac{1}{n}\sum_{i=1}^{n}V_{i}\cdot
\widehat{w}_{g,i}^{treat}\cdot\dot{m}_{g,t,\delta}^{nev}\left(  W_{i}%
;\bar{\beta}_{g,t,\delta}^{nev}\right) \\
&  -\mathbb{E}_{n}\left[  \widehat{w}_{g}^{treat}\cdot\left(  Y_{t}%
-Y_{g-\delta-1}-m_{g,t,\delta}^{nev}\left(  \widehat{\beta}_{g,t,\delta}%
^{nev}\right)  \right)  \right]  \frac{1}{\sqrt{n}}\sum_{i=1}^{n}V_{i}%
\cdot\widehat{w}_{g,i}^{treat}\\
&  =\widehat{I}_{g,t,\delta}^{1,treat}-\widehat{I}_{g,t,\delta}^{2,treat}%
-\widehat{I}_{g,t,\delta}^{3,treat},
\end{align*}
where $\bar{\beta}_{g,t,\delta}^{nev}$ is an intermediate point that satisfies
$\left\vert \bar{\beta}_{g,t,\delta}^{nev}-\beta_{g,t,\delta}^{\ast
,nev}\right\vert \leq\left\vert \widehat{\beta}_{g,t,\delta}^{nev}%
-\beta_{g,t,\delta}^{\ast,nev}\right\vert $ a.s.. From the strong law of large
numbers and the fact that $V$ is mean zero, independent of $W$, it follows
that
\begin{align*}
\widehat{I}_{g,t,\delta}^{1,treat}  &  =\frac{1}{\sqrt{n}}\sum_{i=1}^{n}%
V_{i}\cdot w_{g,i}^{treat}\cdot\left(  Y_{t}-Y_{g-\delta-1}-m_{g,t,\delta
}^{nev}\left(  W_{i};\beta_{g,t,\delta}^{\ast,nev}\right)  \right)
+o_{p^{\ast}}\left(  1\right) \\
\widehat{I}_{g,t,\delta}^{3,treat}  &  =\frac{1}{\sqrt{n}}~\sum_{i=1}^{n}%
V_{i}\cdot w_{g,i}^{treat}\cdot\mathbb{E}\left[  w_{g}^{treat}\cdot\left(
Y_{t}-Y_{g-\delta-1}-m_{g,t,\delta}^{nev}\left(  \beta_{g,t,\delta}^{\ast
,nev}\right)  \right)  \right]  +o_{p^{\ast}}\left(  1\right)
\end{align*}
Similarly, from Assumptions \ref{ass:parametric} and \ref{ass:integrabilitity}%
, and the strong law of large numbers, we conclude that $\widehat{I}%
_{g,t,\delta}^{2,treat}=o_{p^{\ast}}\left(  1\right)  $. Now
(\ref{boot:treated}) follows from combining these results. }

{\small Next, we show%
\begin{equation}
\frac{1}{\sqrt{n}}\sum_{i=1}^{n}V_{i}\cdot\left(  \widehat{\psi}_{g,t,\delta
}^{comp,nev}(W_{i};\widehat{\pi}_{g},\widehat{\beta}_{g,t,\delta}^{nev}%
)-\psi_{g,t,\delta}^{comp,nev}(W_{i};\pi_{g}^{\ast},\beta_{g,t,\delta}%
^{\ast,nev})\right)  =o_{p^{\ast}}\left(  1\right)  . \label{boot:comp}%
\end{equation}
Again, by the mean value theorem, we write
\begin{align*}
&  \frac{1}{\sqrt{n}}\sum_{i=1}^{n}V_{i}\cdot\widehat{\psi}_{g,t,\delta
}^{comp,nev}(W_{i};\widehat{\pi}_{g},\widehat{\beta}_{g,t,\delta}^{nev})\\
&  =\frac{1}{\sqrt{n}}\sum_{i=1}^{n}V_{i}\cdot\widehat{w}_{g}^{comp,nev}%
\left(  W_{i};\widehat{\pi}_{g}\right)  \cdot\left(  Y_{i,t}-Y_{i,g-\delta
-1}-m_{g,t,\delta}^{nev}\left(  W_{i};\beta_{g,t,\delta}^{\ast,nev}\right)
\right) \\
&  -\sqrt{n}\left(  \widehat{\beta}_{g,t,\delta}^{nev}-\beta_{g,t,\delta
}^{\ast,nev}\right)  ^{\prime}\frac{1}{n}\sum_{i=1}^{n}V_{i}\cdot
\widehat{w}_{g}^{comp,nev}\left(  W_{i};\widehat{\pi}_{g}\right)  \cdot\dot
{m}_{g,t,\delta}^{nev}\left(  W_{i};\bar{\beta}_{g,t,\delta}^{nev}\right) \\
&  -\mathbb{E}_{n}\left[  \widehat{w}_{g}^{comp,nev}\left(  \widehat{\pi}%
_{g}\right)  \cdot\left(  Y_{t}-Y_{g-\delta-1}-m_{g,t,\delta}^{nev}\left(
\widehat{\beta}_{g,t,\delta}^{nev}\right)  \right)  \right]  \frac{1}{\sqrt
{n}}\sum_{i=1}^{n}V_{i}\cdot\widehat{w}_{g}^{comp,nev}\left(  W_{i}%
;\widehat{\pi}_{g}\right) \\
&  =\frac{1}{\sqrt{n}}\sum_{i=1}^{n}V_{i}\cdot\frac{\dfrac{p_{g}\left(
X_{i};\pi_{g}^{\ast}\right)  C_{i}}{1-p_{g}\left(  X;\pi_{g}^{\ast}\right)  }%
}{\mathbb{E}_{n}\left[  \dfrac{\widehat{p}_{g}\left(  X;\widehat{\pi}%
_{g}\right)  C}{1-\widehat{p}_{g}\left(  X;\widehat{\pi}_{g}\right)  }\right]
}\cdot\left(  Y_{i,t}-Y_{i,g-\delta-1}-m_{g,t,\delta}^{nev}\left(  W_{i}%
;\beta_{g,t,\delta}^{\ast,nev}\right)  \right) \\
&  +\sqrt{n}\left(  \widehat{\pi}_{g}-\pi_{g}^{\ast}\right)  ^{\prime}\frac
{1}{n}\sum_{i=1}^{n}V_{i}\cdot\frac{\dfrac{C_{i}}{\left(  1-p_{g}\left(
X_{i};\bar{\pi}_{g}\right)  \right)  ^{2}}}{\mathbb{E}_{n}\left[
\dfrac{\widehat{p}_{g}\left(  X;\widehat{\pi}_{g}\right)  C}{1-\widehat{p}%
_{g}\left(  X;\widehat{\pi}_{g}\right)  }\right]  }\cdot\left(  Y_{i,t}%
-Y_{i,g-\delta-1}-m_{g,t,\delta}^{nev}\left(  W_{i};\beta_{g,t,\delta}%
^{\ast,nev}\right)  \right)  \cdot\dot{p}_{g}\left(  X_{i};\bar{\pi}%
_{g}\right) \\
&  -\sqrt{n}\left(  \widehat{\beta}_{g,t,\delta}^{nev}-\beta_{g,t,\delta
}^{\ast,nev}\right)  ^{\prime}\frac{1}{n}\sum_{i=1}^{n}V_{i}\cdot
\widehat{w}_{g}^{comp,nev}\left(  W_{i};\widehat{\pi}_{g}\right)  \cdot\dot
{m}_{g,t,\delta}^{nev}\left(  W_{i};\bar{\beta}_{g,t,\delta}^{nev}\right) \\
&  -\widehat{M}_{g,t,\delta}^{comp}\frac{1}{\sqrt{n}}\sum_{i=1}^{n}V_{i}%
\cdot\frac{\dfrac{p_{g}\left(  X_{i};\pi_{g}^{\ast}\right)  C_{i}}%
{1-p_{g}\left(  X_{i};\pi_{g}^{\ast}\right)  }}{\mathbb{E}_{n}\left[
\dfrac{\widehat{p}_{g}\left(  X;\widehat{\pi}_{g}\right)  C}{1-\widehat{p}%
_{g}\left(  X;\widehat{\pi}_{g}\right)  }\right]  }\\
&  -\widehat{M}_{g,t,\delta}^{comp}\sqrt{n}\left(  \widehat{\pi}_{g}-\pi
_{g}^{\ast}\right)  ^{\prime}\frac{1}{n}\sum_{i=1}^{n}V_{i}\cdot\frac
{\dfrac{C_{i}}{\left(  1-p_{g}\left(  X_{i};\bar{\pi}_{g}\right)  \right)
^{2}}}{\mathbb{E}_{n}\left[  \dfrac{\widehat{p}_{g}\left(  X;\widehat{\pi}%
_{g}\right)  C}{1-\widehat{p}_{g}\left(  X;\widehat{\pi}_{g}\right)  }\right]
}\cdot\dot{p}_{g}\left(  X_{i};\bar{\pi}_{g}\right) \\
&  =\widehat{I}_{g,t,\delta}^{1,comp}+\widehat{I}_{g,t,\delta}^{2,comp}%
-\widehat{I}_{g,t,\delta}^{3,comp}-\widehat{I}_{g,t,\delta}^{4,comp}%
-\widehat{I}_{g,t,\delta}^{5,comp},
\end{align*}
where $\bar{\beta}_{g,t,\delta}^{nev}$ and $\bar{\pi}_{g}$ are intermediate
points that satisfies $\left\vert \bar{\beta}_{g,t,\delta}^{nev}%
-\beta_{g,t,\delta}^{\ast,nev}\right\vert \leq\left\vert \widehat{\beta
}_{g,t,\delta}^{nev}-\beta_{g,t,\delta}^{\ast,nev}\right\vert $ a.s. and
$\left\vert \bar{\pi}_{g}-\pi_{g}^{\ast}\right\vert \leq\left\vert
\widehat{\pi}_{g}-\pi_{g}^{\ast}\right\vert $ a.s., respectively, and
\[
\widehat{M}_{g,t,\delta}^{comp}=\mathbb{E}_{n}\left[  \widehat{w}%
_{g}^{comp,nev}\left(  \widehat{\pi}_{g}\right)  \cdot\left(  Y_{t}%
-Y_{g-\delta-1}-m_{g,t,\delta}^{nev}\left(  \widehat{\beta}_{g,t,\delta}%
^{nev}\right)  \right)  \right]  .
\]
From the strong law of large numbers and the fact that $V$ is mean zero, has
variance one, and is independent of $W$, it follows that
\begin{align*}
\widehat{I}_{g,t,\delta}^{1,comp}  &  =\frac{1}{\sqrt{n}}\sum_{i=1}^{n}%
V_{i}\cdot w_{g}^{comp,nev}\left(  W_{i};\pi_{g}^{\ast}\right)  \cdot\left(
Y_{i,t}-Y_{i,g-\delta-1}-m_{g,t,\delta}^{nev}\left(  W_{i};\beta_{g,t,\delta
}^{\ast,nev}\right)  \right)  +o_{p^{\ast}}\left(  1\right)  ,\\
\widehat{I}_{g,t,\delta}^{3,treat}  &  =\frac{1}{\sqrt{n}}\sum_{i=1}^{n}%
V_{i}\cdot w_{g}^{comp,nev}\left(  W_{i};\pi_{g}^{\ast}\right)  \cdot
\mathbb{E}\left[  w_{g}^{comp}\cdot\left(  Y_{t}-Y_{g-\delta-1}-m_{g,t,\delta
}^{nev}\left(  \beta_{g,t,\delta}^{\ast,nev}\right)  \right)  \right]
+o_{p^{\ast}}\left(  1\right)  .
\end{align*}
Similarly, from Assumptions \ref{ass:parametric} and \ref{ass:integrabilitity}%
, and the strong law of large numbers, we conclude that%
\[
\widehat{I}_{g,t,\delta}^{2,comp}=\widehat{I}_{g,t,\delta}^{4,comp}%
=\widehat{I}_{g,t,\delta}^{5,comp}=o_{p^{\ast}}\left(  1\right)  .
\]
Now (\ref{boot:comp}) follows from combining these results. }

{\small Next, we show that%
\begin{equation}
\frac{1}{\sqrt{n}}\sum_{i=1}^{n}V_{i}\cdot\left(  \widehat{\psi}%
_{g,t}^{est,nev}(W_{i};\widehat{\pi}_{g},\widehat{\beta}_{g,t,\delta}%
^{nev})-\psi_{g,t}^{est,nev}(W_{i};\pi_{g}^{\ast},\beta_{g,t,\delta}%
^{\ast,nev})\right)  =o_{p^{\ast}}\left(  1\right)  . \label{boot:est}%
\end{equation}
From the strong law of large numbers and Assumptions \ref{ass:parametric} and
\ref{ass:integrabilitity},%
\begin{align*}
&  \frac{1}{\sqrt{n}}\sum_{i=1}^{n}V_{i}\cdot\widehat{\psi}_{g,t}%
^{est,nev}(W_{i};\widehat{\pi}_{g},\widehat{\beta}_{g,t,\delta}^{nev})\\
&  =\frac{1}{\sqrt{n}}\sum_{i=1}^{n}V_{i}\cdot\left(  l_{g,t}^{or,nev}\left(
W_{i};\widehat{\beta}_{g,t,\delta}^{nev}\right)  ^{\prime}\cdot M_{g,t,\delta
}^{dr,nev,1}+l_{g}^{ps,nev}\left(  W_{i};\widehat{\pi}_{g}\right)  ^{\prime
}\cdot M_{g,t,\delta}^{dr,nev,2}\right)  +o_{p^{\ast}}\left(  1\right) \\
&  =\frac{1}{\sqrt{n}}\sum_{i=1}^{n}V_{i}\cdot l_{g,t}^{est}\left(
W_{i};\widehat{\kappa}_{g,t}^{nev}\right)  +o_{p^{\ast}}\left(  1\right)  ,
\end{align*}
where, for a generic $\kappa_{g,t}^{nev}=\left(  \pi_{g}^{\prime}%
,\beta_{g,t,\delta}^{nev^{\prime}}\right)  ^{\prime},$
\[
l_{g,t}^{est}\left(  W;\kappa_{g,t}^{nev}\right)  =l_{g,t}^{or,nev}\left(
W;\beta_{g,t,\delta}^{nev}\right)  ^{\prime}\cdot M_{g,t,\delta}%
^{dr,nev,1}+l_{g}^{ps,nev}\left(  W_{i};\pi_{g}\right)  ^{\prime}\cdot
M_{g,t,\delta}^{dr,nev,2}%
\]
Thus, from Lemma 4.3 in \cite{Newey1994c} and Assumption \ref{ass:parametric},
it follows that
\begin{align*}
&  Var^{\ast}\left(  \frac{1}{\sqrt{n}}\sum_{i=1}^{n}V_{i}\cdot\left(
l_{g,t}^{est}\left(  W_{i};\widehat{\kappa}_{g,t,\delta}^{nev}\right)
-l_{g,t}^{est}\left(  W_{i};\kappa_{g,t,\delta}^{\ast,nev}\right)  \right)
\right) \\
&  =\frac{1}{n}\sum_{i=1}^{n}\left(  l_{g,t}^{est}\left(  W_{i}%
;\widehat{\kappa}_{g,t,\delta}^{nev}\right)  -l_{g,t}^{est}\left(
W_{i};\kappa_{g,t,\delta}^{\ast,nev}\right)  \right)  ^{2}\\
&  =o_{p}\left(  1\right)  ,
\end{align*}
which, in turn, implies (\ref{boot:est}). }

{\small Taking (\ref{boot:treated}), (\ref{boot:comp}),\ and (\ref{boot:est})
together, we then establish (\ref{boot:needed}). Thus, by (\ref{eq:boot.0}),
we have%
\[
\sqrt{n}\left(  \widehat{ATT}_{t\geq\left(  g-\delta\right)  }^{\ast
,dr,nev}-\widehat{ATT}_{t\geq\left(  g-\delta\right)  }^{dr,nev}\right)
\underset{\ast}{\overset{d}{\rightarrow}}N(0,\Sigma).
\]
}

{\small Finally, by the continuous mapping theorem, see e.g. Theorem 10.8 in
\cite{kosorok2008}, for any continuous functional $\Gamma(\cdot)$
\[
\Gamma\left(  \sqrt{n}\left(  \widehat{ATT}_{t\geq\left(  g-\delta\right)
}^{\ast,dr,nev}-\widehat{ATT}_{t\geq\left(  g-\delta\right)  }^{dr,nev}%
\right)  \right)  \underset{\ast}{\overset{d}{\rightarrow}}\Gamma\left(
N(0,\Sigma)\right)  ,
\]
concluding our proof. $\square$ }

\section*{Appendix B: Additional Results for Repeated Cross
Sections\label{App.B}}

\renewcommand{\thesection}{B}
\renewcommand{\theassumption}{B.\arabic{assumption}}
\renewcommand{\thelemma}{B.\arabic{lemma}} \setcounter{lemma}{0}
\setcounter{equation}{0} \setcounter{figure}{0} \setcounter{section}{0}
\renewcommand{\thetheorem}{B.\arabic{theorem}} \setcounter{theorem}{0}
\setcounter{assumption}{0} {\small In this section we extend our
identification results to the case where one has access to repeated cross
sections data instead of panel data. Here we assume that for each unit in the
pooled sample, we observe $(Y,G_{2},\ldots,G_{\mathcal{T}},C,T,X)$ where
$T\in\{1,\ldots,\mathcal{T}\}$ denotes the time period when that unit is
observed. Let $T_{t}=1$ if an observation is observed at time $t$, and zero
otherwise. }

{\small We assume that random samples are available for each time period. }

\begin{assumption}
{\small \label{sampling_rcs}Conditional of $T=t$, the data are independent and
identically distributed from the distribution of $\left(  Y_{t},G_{2}%
,\ldots,G_{\mathcal{T}},C,X\right)  ,$ for all $t=1,\dots,\mathcal{T}$., with
$\left(  G_{2},\ldots,G_{\mathcal{T}},C,X\right)  $ being invariant to $T$. }
\end{assumption}

{\small Assumption \ref{sampling_rcs} implies that our sample consists of
random draws from the mixture distribution
\[
F_{M}(y,g_{2},\ldots,g_{\mathcal{T}},c,t,x)=\sum_{t=1}^{\mathcal{T}}%
\lambda_{t}\cdot F_{Y,G_{2},\ldots,G_{\mathcal{T}},C,X|T}(y,g_{2}%
,\ldots,g_{\mathcal{T}},c,x|t),
\]
where $\lambda_{t}=P(T_{t}=1)$. It also rules-out compositional changes across
time. This assumption is related to the sampling assumption imposed by
\cite{Abadie2005} and \cite{SantAnna2020} in the two periods, two groups DiD 
setup. Notice that, once one conditions on the time period, then expectations
under the mixture distribution correspond to population expectations. Also,
because $X$, $G_{g}$, and $C$ are observed for all units, by the stationarity
condition one can use draws from the mixture distribution to estimate the
generalized propensity score. With some abuse of notation, we then use
$p_{g,s}\left(  X\right)  $ as a short notation for $\mathbb{E}_{M}\left[
G_{g}|X,G_{g}+\left(  1-D_{s}\right)  \left(  1-G_{g}\right)  =1\right]  $,
where $\mathbb{E}_{M}\left[  \cdot\right]  $ denotes expectations with respect
to $F_{M}\left(  \cdot\right)  $. Also, we use $p_{g}(X)=p_{g,\mathcal{T}%
}(X)=\mathbb{E}_{M}\left[  G_{g}|X,G_{g}+C=1\right]  $. }

{\small Before formalizing all the results, we need to introduce some
additional notation. Let $m_{c,t}^{rc,nev}\left(  x\right)  \equiv
\mathbb{E}_{M}[Y|X=x,C=1,T=t]$, $m_{g,t}^{rc,treat}\left(  x\right)
\equiv\mathbb{E}_{M}[Y|X=x,G_{g}=1,T=t]$ and $m_{s,t}^{rc,ny}\left(  x\right)
\equiv\mathbb{E}_{M}[Y|X=x,D_{s}=0,G_{g}=0,T=t]$. Consider the weights%
\begin{align*}
w^{treat}\left(  a,b\right)   &  =\left.  T_{b}\cdot G_{a}\right/
\mathbb{E}_{M}\left[  T_{b}\cdot G_{a}\right]  ,\\
w_{nev}^{comp}\left(  a,b\right)   &  =\left.  \dfrac{T_{b}\cdot p_{a}\left(
X\right)  C}{1-p_{a}\left(  X\right)  }\right/  \mathbb{E}_{M}\left[
\dfrac{T_{b}\cdot p_{a}\left(  X\right)  C}{1-p_{a}\left(  X\right)  }\right]
,\\
w_{ny}^{comp}\left(  a,b,s\right)   &  =\left.  \dfrac{T_{b}\cdot
p_{a,s}\left(  X\right)  \left(  1-D_{b}\right)  \left(  1-G_{a}\right)
}{1-p_{a,s}\left(  X\right)  }\right/  \mathbb{E}_{M}\left[  \dfrac{T_{b}\cdot
p_{a,s}\left(  X\right)  \left(  1-D_{b}\right)  \left(  1-G_{a}\right)
}{1-p_{a,s}\left(  X\right)  }\right]  .
\end{align*}
Finally, consider the outcome regression (OR) estimands,%
\begin{align*}
ATT_{or,rc}^{nev}\left(  g,t;\delta\right)   &  =\mathbb{E}_{M}\left[
\frac{G_{g}}{\mathbb{E}_{M}\left[  G_{g}\right]  }\left(  \left(
m_{g,t}^{rc,treat}\left(  X\right)  -m_{g,g-\delta-1}^{rc,treat}\left(
X\right)  \right)  -\left(  m_{c,t}^{rc,nev}\left(  X\right)  -m_{c,g-\delta
-1}^{rc,nev}\left(  X\right)  \right)  \right)  \right]  ,\\
ATT_{or,rc}^{ny}\left(  g,t;\delta\right)   &  =\mathbb{E}_{M}\left[
\frac{G_{g}}{\mathbb{E}_{M}\left[  G_{g}\right]  }\left(  \left(
m_{g,t}^{rc,treat}\left(  X\right)  -m_{g,g-\delta-1}^{rc,treat}\left(
X\right)  \right)  -\left(  m_{t+\delta,t}^{rc,ny}\left(  X\right)
-m_{t+\delta,g-\delta-1}^{rc,ny}\left(  X\right)  \right)  \right)  \right]  ,
\end{align*}
the inverse probability weighted (IPW) estimands%
\begin{align*}
ATT_{ipw,rc}^{nev}\left(  g,t;\delta\right)   &  =\mathbb{E}_{M}\left[
\left(  w^{treat}\left(  g,t\right)  -w^{treat}\left(  g,g-\delta-1\right)
\right)  \cdot Y\right] \\
&  -\mathbb{E}_{M}\left[  \left(  w_{nev}^{comp}\left(  g,t\right)
-w_{nev}^{comp}\left(  g,g-\delta-1\right)  \right)  \cdot Y\right]  ,\\
ATT_{ipw,rc}^{ny}\left(  g,t;\delta\right)   &  =\mathbb{E}_{M}\left[  \left(
w^{treat}\left(  g,t\right)  -w^{treat}\left(  g,g-\delta-1\right)  \right)
\cdot Y\right] \\
&  -\mathbb{E}_{M}\left[  \left(  w_{ny}^{comp}\left(  g,t,t+\delta\right)
-w_{ny}^{comp}\left(  g,g-\delta-1,t+\delta\right)  \right)  \cdot Y\right]  ,
\end{align*}
and the doubly-robust (DR) estimands%
\begin{align*}
ATT_{dr,rc}^{nev}\left(  g,t;\delta\right)   &  =\mathbb{E}_{M}\left[
\frac{G_{g}}{\mathbb{E}_{M}\left[  G_{g}\right]  }\left(  \left(
m_{g,t}^{rc,treat}\left(  X\right)  -m_{g,g-\delta-1}^{rc,treat}\left(
X\right)  \right)  -\left(  m_{c,t}^{rc,nev}\left(  X\right)  -m_{c,g-\delta
-1}^{rc,nev}\left(  X\right)  \right)  \right)  \right] \\
&  +\mathbb{E}_{M}\left[  w^{treat}\left(  g,t\right)  \left(  Y-m_{g,t}%
^{rc,treat}\left(  X\right)  \right)  -w^{treat}\left(  g,g-\delta-1\right)
\left(  Y-m_{g,g-\delta-1}^{rc,treat}\left(  X\right)  \right)  \right] \\
&  -\mathbb{E}_{M}\left[  w_{nev}^{comp}\left(  g,t\right)  \left(
Y-m_{c,t}^{rc,nev}\left(  X\right)  \right)  -w_{nev}^{comp}\left(
g,g-\delta-1\right)  \left(  Y-m_{c,g-\delta-1}^{rc,nev}\left(  X\right)
\right)  \right]  ,\\
ATT_{dr,rc}^{ny}\left(  g,t;\delta\right)   &  =\mathbb{E}_{M}\left[
\frac{G_{g}}{\mathbb{E}_{M}\left[  G_{g}\right]  }\left(  \left(
m_{g,t}^{rc,treat}\left(  X\right)  -m_{g,g-\delta-1}^{rc,treat}\left(
X\right)  \right)  -\left(  m_{t+\delta,t}^{rc,ny}\left(  X\right)
-m_{t+\delta,g-\delta-1}^{rc,ny}\left(  X\right)  \right)  \right)  \right] \\
&  +\mathbb{E}_{M}\left[  w^{treat}\left(  g,t\right)  \left(  Y-m_{g,t}%
^{rc,treat}\left(  X\right)  \right)  -w^{treat}\left(  g,g-\delta-1\right)
\left(  Y-m_{g,g-\delta-1}^{rc,treat}\left(  X\right)  \right)  \right] \\
&  -\mathbb{E}_{M}\left[  w_{ny}^{comp}\left(  g,t,t+\delta\right)  \left(
Y-m_{t+\delta,t}^{rc,ny}\left(  X\right)  \right)  -w_{ny}^{comp}\left(
g,g-\delta-1,t+\delta\right)  \left(  Y-m_{t+\delta,g-\delta-1}^{rc,ny}\left(
X\right)  \right)  \right]  .
\end{align*}
}

{\small The OR, IPW and DR estimands respectively generalize
\cite{Heckman1997}, \cite{Abadie2005} and \cite{SantAnna2020} estimands for
the two groups, two periods DiD setup to the staggered DiD setup with multiple
periods and multiple groups. }

\begin{theorem}
{\small \label{thm:att_rcs} Let Assumptions \ref{ass:irrev},
\ref{ass:anticipation}, \ref{ass:overlap}, and \ref{sampling_rcs} hold. }

{\small $\left(  i\right)  $ If Assumption \ref{ass:common-trends} in the main
text holds, then, for all $g$ and $t$ such that $g\in\mathcal{G}_{\delta}$,
$t\in\left\{  2,\dots\mathcal{T-\delta}\right\}  $ and $t\geq g-\delta$,%
\[
ATT\left(  g,t\right)  =ATT_{ipw,rc}^{nev}\left(  g,t;\delta\right)
=ATT_{or,rc}^{nev}\left(  g,t;\delta\right)  =ATT_{dr,rc}^{nev}\left(
g,t;\delta\right)  .
\]
}

{\small $\left(  ii\right)  $ If Assumption \ref{ass:common-trends-ny} in the
main text holds, then, for all $g$ and $t$ such that $g\in\mathcal{G}_{\delta
}$, $t\in\left\{  2,\dots\mathcal{T-\delta}\right\}  $ and $t\geq g-\delta$,%
\[
ATT\left(  g,t\right)  =ATT_{ipw,rc}^{ny}\left(  g,t;\delta\right)
=ATT_{or,rc}^{ny}\left(  g,t;\delta\right)  =ATT_{dr,rc}^{ny}\left(
g,t;\delta\right)  .
\]
}
\end{theorem}

{\small We defer the proof of Theorem \ref{thm:att_rcs} to the Supplementary
Appendix. The identification results in Theorem \ref{thm:att_rcs} suggest a
simple two-step estimation procedure for the $ATT(g,t)$ with repeated
cross-section data that is analogous to the panel data case discussed in
Section \ref{sec:estimation}. The asymptotic properties of such two-step
estimators follow from analogous arguments; the details are omitted for
brevity. Likewise, we can aggregate these estimators to provide summary
measures of the causal effects like those discussed in Section \ref{sec:4}. }

\onehalfspacing{\small
\bibliographystyle{jasa}
\bibliography{DID}
}
\end{document}